\documentclass[12pt]{article}

\usepackage{amsmath}
\usepackage{amssymb}
\usepackage{cite}
\usepackage{bm}
\usepackage{epsfig}
\usepackage{color}
\usepackage{here}

\makeatletter
\numberwithin{equation}{section}

\DeclareMathOperator{\card}{card}

\newcommand{\h}{{\cal H}}

\makeatletter
\DeclareRobustCommand\widecheck[1]{{\mathpalette\@widecheck{#1}}}
\def\@widecheck#1#2{
    \setbox\z@\hbox{\m@th$#1#2$}
    \setbox\tw@\hbox{\m@th$#1
       \widehat{
          \vrule\@width\z@\@height\ht\z@
          \vrule\@height\z@\@width\wd\z@}$}
    \dp\tw@-\ht\z@
    \@tempdima\ht\z@ \advance\@tempdima2\ht\tw@ \divide\@tempdima\thr@@
    \setbox\tw@\hbox{%
       \raise\@tempdima\hbox{\scalebox{1}[-1]{\lower\@tempdima\box
\tw@}}}
    {\ooalign{\box\tw@ \cr \box\z@}}}
\makeatother

\newcounter{aff}

\makeatother

\begin{document}
\begin{titlepage}
\begin{flushright}
{\footnotesize OCU-PHYS 495, NITEP 5}
\end{flushright}
\begin{center}
{\Large\bf
ABJM Matrix Model and 2D Toda Lattice Hierarchy}\\
\bigskip\bigskip
{\large Tomohiro Furukawa\,\footnote{\tt furukawa@sci.osaka-cu.ac.jp}
\quad and \quad
Sanefumi Moriyama\,\footnote{\tt moriyama@sci.osaka-cu.ac.jp}}\\
\bigskip
${}^{*\dagger}$\,{\it Department of Physics, Graduate School of Science, Osaka City University,}\\
{\it Sumiyoshi-ku, Osaka 558-8585, Japan}\\[3pt]
${}^\dagger$\,{\it Osaka City University Advanced Mathematical Institute (OCAMI),}\\
{\it Sumiyoshi-ku, Osaka 558-8585, Japan}\\[3pt]
${}^\dagger$\,{\it Nambu Yoichiro Institute of Theoretical and Experimental Physics (NITEP),}\\
{\it Sumiyoshi-ku, Osaka 558-8585, Japan}
\end{center}

\begin{abstract}
It was known that one-point functions in the ABJM matrix model (obtained by applying the localization technique to one-point functions of the half-BPS Wilson loop operator in the ABJM theory) satisfy the Jacobi-Trudi formula, which strongly indicates the integrable structure of the system.
In this paper, we identify the integrable structure of two-point functions in the ABJM matrix model as the two-dimensional Toda lattice hierarchy.
The identification implies infinitely many non-linear differential equations for the generating function of the two-point functions.
\end{abstract}

\end{titlepage}

\setcounter{footnote}{0}

\tableofcontents

\section{Introduction}

The study of solitary waves starts from an important observation by J.~Scott Russell in 1834 in a channel in Scotland.
After the non-linear differential equation for the solitary waves was established, a recurrence phenomenon for the solutions (solitons) was observed \cite{ZK} though it was mysterious why the solutions are stable.
Interestingly, it was found that there are an infinite number of non-linear differential equations compatible with the original equation, which serve as infinite conserved charges.
The whole set of infinite non-linear differential equations was called integrable hierarchy.
This structure was generalized to many systems, where the two-dimensional Toda lattice (2DTL) system is one of the ultimate ones.
In these integrable hierarchies the structure of an infinite-dimensional Grassmann manifold was found and the findings were further interpreted as a hidden gl$(\infty)$ symmetry for the infinite number of equations and conserved charges by M.~Sato and the Kyoto school \cite{S,DJKM,MJD}.
The Sato theory claims that, when the most general soliton solution (tau function) is expanded with the Schur functions, the coefficients satisfy the Pl\"ucker relations and the whole set of the relations is equivalent to the integrable hierarchy.
Several famous Pl\"ucker relations are known as the Giambelli formula or the Jacobi-Trudi formula.
Hence, the appearance of these formulas strongly implies a hidden integrability of the system.
The integrable structure is beautifully encoded in a free fermion system \cite{DJKM,JM,MJD}.

Aside from the progress in the non-linear differential equations, the study of non-perturbative effects in string theory is one of the central problems in modern particle physics.
It was found that ten-dimensional perturbative string theory extends non-perturbatively to eleven-dimensional M-theory where membranes (M2-branes) are fundamental excitations.
The worldvolume theory of the M2-branes was mysterious for a long time and finally it was proposed \cite{ABJM,HLLLP2,ABJ} that it is the three-dimensional ${\cal N}=6$ superconformal Chern-Simons theory that describes the M2-branes.
The partition function and one-point functions of the half-BPS Wilson loop $W_\lambda$ labeled by a Young diagram $\lambda$ \cite{MPtop,DT} (which we loosely call $\langle 1\rangle$ and $\langle W_\lambda\rangle$ respectively in this introduction and postpone the explanation to the next section) are originally defined with infinite-dimensional path integrals.
Using the localization technique \cite{KWY}, however, these correlation functions on $S^3$ reduce to finite-dimensional multiple integrals (ABJM matrix model).
Although the derivation from the localization technique is missing, a corresponding multiple integral expression for two-point functions $\langle W_\lambda\overline W_\mu\rangle$ was proposed and studied carefully in \cite{KM}.

In the study of the matrix model, it was found that the partition function $\langle 1\rangle$ and the one-point function $\langle W_\lambda\rangle$ are expressed as a determinant, similar to the Giambelli formula though associated with an extra shift \cite{MatsumotoM}, which we call {\it shifted Giambelli relation}.
The shifted Giambelli relation is organized very orderly, so that we can even take it as a principle.
In fact, aside from the physical arguments given in \cite{KM}, one of the reasons we consider that the proposal on the multiple integral expression for the two-point function $\langle W_\lambda\overline W_\mu\rangle$ is appropriate is because the expression also enjoys the shifted Giambelli relation \cite{KM}.

Using this relation, subsequently we can prove the original Giambelli formula \cite{HHMO,MatsunoM} and the Jacobi-Trudi formula \cite{FM}.
As mentioned above, these formulas are famous in the context of solvable systems.
In fact, in the study of soliton systems or solvable lattice systems, similar formulas appear (see, for example, \cite{KOS,HL,AKLTZ,AZ}).
Also, in \cite{md,md9} in a series of variations of the Schur function, as a final one called the ninth variation, the Jacobi-Trudi formula was regarded as the definition of the Schur function and many interesting properties such as the Giambelli formula can still be derived from this simple setup.
Hence, the proof of the Giambelli formula or the Jacobi-Trudi formula in the ABJM matrix model strongly suggests the integrable structure in behind.
It is then interesting to ask what soliton equations the generating function of correlation functions in the ABJM matrix model follows.

In this paper, to clarify the integrable hierarchy structure in the ABJM matrix model, we utilize the fermionic construction \cite{DJKM,JM,MJD,AKLTZ,AZ}.
We prove that the two-point functions in the ABJM matrix model and the 2DTL system share exactly the same integrable structure.

Let us summarize several implications of our results for pure string theorists, who may not be very interested in the integrable structure itself.
Our results indicate that the generating function $F_{n}(\cdots,t_{-1},t_1,\cdots)$ of the two-point functions in the ABJM matrix model $\langle W_\lambda\overline W_\mu\rangle$,
\begin{align}
F_{n}(\cdots,t_{-1},t_1,\cdots)
=\sum_{\lambda,\mu}\langle W_\lambda\overline{W}_\mu\rangle(-n)
\widetilde{s}_\lambda(t_1,\cdots)\widetilde{s}_\mu(-t_{-1},\cdots),
\label{generating}
\end{align}
(with $\widetilde{s}_\lambda(t_1,\cdots)$ being the Schur function and $n=-M$ being a parameter of the two-point function, see section \ref{abjm}) satisfies infinitely many non-linear differential equations where the first one is known as the 2DTL equation,
\begin{align}
\frac{\partial^2 u_n}{\partial t_1 \partial t_{-1}}=e^{u_n-u_{n-1}}-e^{u_{n+1}-u_n},
\label{2dtleq}
\end{align}
with $u_n=\ln(F_{n+1}/F_n)$ (see section \ref{correspond} for more clarifications).
A physical interpretation for these differential equations awaits to be found.
If the reader is only interested in the partition function $\langle 1\rangle$ or the one-point functions of the half-BPS Wilson loop $\langle W_\lambda\rangle$ derived with sound arguments of the localization technique, by simple reduction, we find immediately that the generating function of the one-point functions follows the modified Kadomtsev-Petviashvili (mKP) hierarchy.
Nevertheless, we stress that the study in the framework of the two-point functions is important for the identification.

We also derive a Pl\"ucker relation that the two-point functions satisfy, which is a natural generalization of the Giambelli formula since it reduces to the usual one \cite{MatsunoM} when we restrict one of the representations to be trivial.
Due to this reason, we call this relation the Giambelli formula for the two-point functions.

The organization of this paper is as follows.
In the next section, we review the ABJM matrix model and its determinant representation indicating the integrable structure, as well as the fermionic construction useful for studying integrable systems.
After the preparations, in the subsequent section, we propose a correspondence between the two-point functions in the ABJM matrix model and the 2DTL integrable hierarchy.
In appendix A we prove a generalization of the Wick's theorem which is useful for the identification of the correspondence.
In appendix B we derive the Giambelli formula for the two-point functions.

\section{ABJM matrix model and fermionic construction}

In this section we review two contents, the ABJM matrix model and the fermionic construction, especially focusing on the aspect of the integrable structure.

\subsection{Young diagram}

Both the half-BPS Wilson loops in the ABJM theory and the excitations in the fermionic construction are characterized by Young diagrams, which have several major notations.
Before proceeding to the reviews of the two contents, we start with confirming notations for the Young diagram $\lambda$.

In the first notation, we simply list the number of horizontal boxes or vertical boxes as
\begin{align}
\lambda=[\lambda_1,\lambda_2,\cdots]=[\lambda'_1,\lambda'_2,\cdots]'.
\end{align}
Namely, $\lambda_i\in{\mathbb Z}_{\ge 0}$ is the box number in the $i$-th row of the Young diagram appearing in the decreasing order, $\lambda_1\ge\lambda_2\ge\cdots$.
Also, $\lambda'$ denotes the transpose of the Young diagram $\lambda$ and hence $\lambda'_j$ is the box number in the $j$-th column of the Young diagram, $\lambda'_1\ge\lambda'_2\ge\cdots$.
We only consider the Young diagram with its total box number being finite, which implies that the number of non-zero $\lambda_i$ or $\lambda'_j$ is finite, though it is sometimes considered to be followed by infinite series of zeros.

The second one is the standard Frobenius notation, where the Young diagram is expressed by listing the arm lengths and the leg lengths as
\begin{align}
\lambda=(\alpha_1,\alpha_2,\cdots,\alpha_r|\beta_1,\beta_2,\cdots,\beta_r).
\end{align}
Here the arm length $\alpha_i\in{\mathbb Z}_{\ge 0}$ is defined as the $i$-th number of horizontal boxes counted from the diagonal box, while the leg length $\beta_j\in{\mathbb Z}_{\ge 0}$ is the $j$-th number of vertical boxes,
\begin{align}
\alpha_i=\lambda_i-i,\quad
\beta_j=\lambda'_j-j,\quad
r=\card\{i|\alpha_i\ge 0\}=\card\{j|\beta_j\ge 0\},
\end{align}
both of which appear in the strictly decreasing order, $\alpha_1>\alpha_2>\cdots>\alpha_r\ge 0$ and $\beta_1>\beta_2>\cdots>\beta_r\ge 0$.
Here we count the number of boxes from the right or the bottom of the diagonal box.

Sometimes we consider a pair of Young diagrams $\lambda\overline\mu=(\lambda,\mu)$ as a whole and call it the composite Young diagram \cite{Moens}.
Besides the Frobenius notation for $\lambda$, we can introduce a Frobenius notation for $\overline\mu$ using negative lengths of arms and legs, so that the composite Young diagram $\lambda\overline\mu$ is expressed with the whole set of integers.
Namely, for the Young diagram $\mu$, we define the negative arm length $\overline{\alpha}_i\in{\mathbb Z}_{<0}$ and the negative leg length $\overline{\beta}_j\in{\mathbb Z}_{<0}$ by
\begin{align}
&\overline{\alpha}_i=-\mu'_i+i-1,\quad
\overline{\beta}_j=-\mu_j+j-1,\quad
\overline{r}=\card\{i|\overline{\alpha}_i\le -1\}=\card\{j|\overline{\beta}_j\le -1\},
\end{align}
satisfying the strictly increasing order, $\overline\alpha_1<\overline\alpha_2<\cdots<\overline\alpha_{\overline{r}}<0$, $\overline\beta_1<\overline\beta_2<\cdots<\overline\beta_{\overline{r}}<0$
and introduce the Frobenius notation for $\overline\mu$ as
$\overline\mu
=(\overline{\alpha}_{\overline{r}},\overline{\alpha}_{\overline{r}-1},\cdots,\overline{\alpha}_{1}|
\overline{\beta}_{\overline{r}},\overline{\beta}_{\overline{r}-1}\cdots,\overline{\beta}_{1})$.
Then we can combine the two sets of integers to express the Frobenius notation for the composite Young diagram $\lambda\overline\mu$ by the disjoint sum, 
\begin{align}
\lambda\overline\mu
=(\overline{\alpha}_{\overline{r}},\overline{\alpha}_{\overline{r}-1},\cdots,\overline{\alpha}_{1},
\alpha_1,\alpha_2,\cdots,\alpha_r|
\overline{\beta}_{\overline{r}},\overline{\beta}_{\overline{r}-1}\cdots,\overline{\beta}_{1},
\beta_1,\beta_2,\cdots,\beta_r).
\label{Fcomp}
\end{align}
This definition of the Frobenius notation for the composite Young diagram by the disjoint sum is generalized beautifully in the next notation.

\begin{figure}[!t]
\centering\includegraphics[scale=0.5,angle=90]{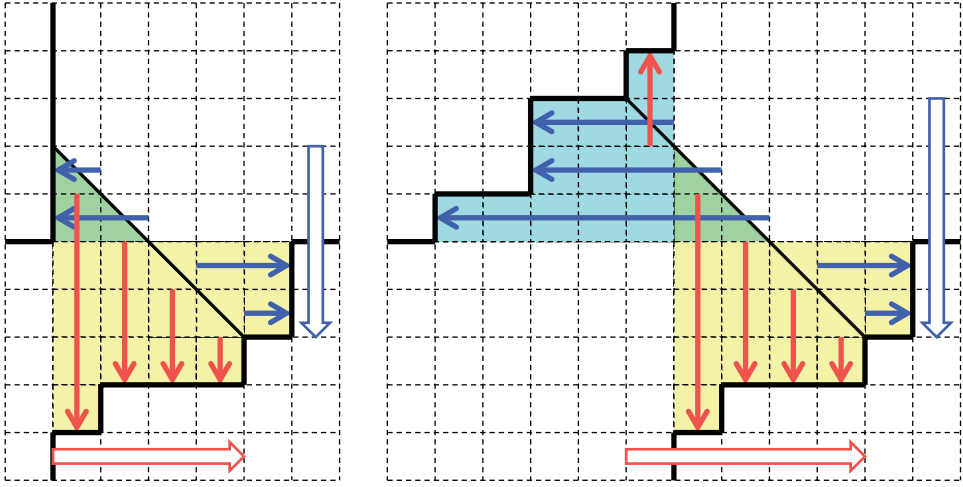}
\caption{Shifted Frobenius notation for $M>0$.
(Left) For a single Young diagram, the $M$-shifted Frobenius notation is given by counting the numbers of horizontal boxes and vertical boxes from the diagonal shifted by $M$.
For the example of $\lambda=[5,5,4,1]$ and $M=2$, the shifted Frobenius notation is given by
$(A_1,A_2|B_1,B_2,B_3,B_4)=(2,1|5,3,2,1)$.
Since there are $M$ more leg lengths than arm lengths, we supplement the $M$-shifted Frobenius notation by $M$ auxiliary arm lengths as $(\overline A_2,\overline A_1,A_1,A_2|B_1,B_2,B_3,B_4)=(-1,-2,2,1|5,3,2,1)$.
We list the numbers in the order of the blank blue arrow and the blank red arrow.
(Right)
The introduction of the auxiliary arm lengths is naturally generalized to a composite Young diagram.
For a composite Young diagram $\lambda\overline\mu$, we consider a pair of the rims and define the $M$-shifted Frobenius notation by the lengths from the shifted diagonal to the rims.
For the example of $\lambda\overline\mu=([5,5,4,1],[4,3,3,1,1])$ and $M=2$, the shifted Frobenius notation is $(\overline A_3,\overline A_2,\overline A_1,A_1,A_2|\overline B_1,B_1,B_2,B_3,B_4)=(-3,-4,-7,2,1|{-2},5,3,2,1)$.
}
\label{compositeY}
\end{figure}

The third one is an extension of the previous Frobenius notation which we call the $M$-shifted Frobenius notation.
We first explain the idea for a single Young diagram and turn to the explicit definition later for a composite Young diagram which may seem clearer.
In this notation (see figure \ref{compositeY}), we shift the diagonal by $M$ to the right if $M$ is a positive integer or by $|M|$ to the left if $M$ is a negative integer.
Then we can define a similar Frobenius notation using the shifted diagonal \cite{MatsumotoM,FM}.
For this shifted Frobenius notation let us use the upper-case Latin characters in the following.
Namely, for a single Young diagram $\lambda$, in the $M\ge 0$ case, the arm length and the leg length are given by $A_i=\lambda_i-i-M$, $B_j=\lambda_j'-j+M$, while, in the $M\le 0$ case, the arm length and the leg length are $A_i=\lambda_i-i+|M|$, $B_j=\lambda_j-j-|M|$.
Apparently the numbers of arm lengths and leg lengths are not equal for $M\ne 0$ since $\card\{i|A_i\ge 0\}=\card\{j|B_j\ge 0\}-M$.
To equate the numbers, it is often convenient to introduce auxiliary arm length $\overline{A}_i=i-1-M$, $(1\le i\le M)$ for $M>0$ and auxiliary leg length $\overline{B}_j=j-1-|M|$, $(1\le j\le|M|)$ for $M<0$ and interpret their absolute values as the lengths to ``the rim'' (see figure \ref{compositeY} again).
Namely, if we interpret the arm/leg lengths $A_i,B_j$ as the box numbers to the rim (the bold black polygonal lines) of the Young diagram in the lower-right part and apply the interpretation to the auxiliary arm/leg lengths $\overline A_i,\overline B_j$ as well, it is natural to extend the rim to the upper-left part by straight lines.
The idea of introducing extra auxiliary lengths may look artificial at first sight.
Using this shifted Frobenius notation, however, as we review in the next subsection, we can define the shifted Giambelli relation and with the relation we can prove the original Giambelli formula \cite{MatsunoM} and the Jacobi-Trudi formula \cite{FM}.

\begin{figure}[!t]
\centering\includegraphics[scale=0.5,angle=-90]{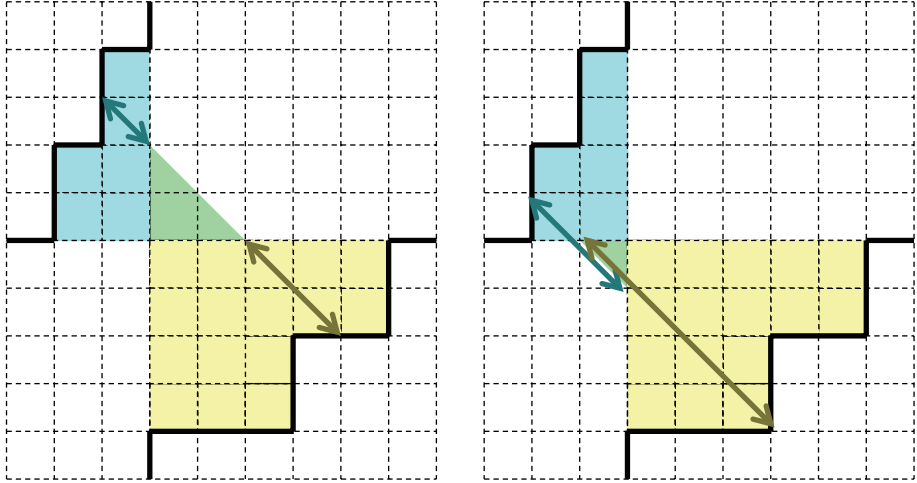}
\caption{A pictorial interpretation of $R$ and $\overline{R}$ for $M$ being positive or negative.
$R$ and $\overline{R}$ are respectively given by the length of the dark yellow arrow and that of the dark cyan arrow.
For the left example we have $(M,R,\overline R)=(2,2,1)$, while for the right one we have $(M,R,\overline R)=(-1,4,2)$.
For the case of $M<0$, $R$ and $\overline{R}$ always satisfy $R\ge|M|$ and $\overline{R}\ge|M|$.
}
\label{RRbar}
\end{figure}

The ideas of the shifted Frobenius notation and the composite Young diagram match well.
Let us define the $M$-shifted Frobenius notation for a composite Young diagram $\lambda\overline\mu$ as follows.
Namely, we define the arm lengths and the leg lengths appearing in the $M$-shifted Frobenius notation for the composite Young diagram $\lambda\overline\mu$
\begin{align}
\lambda\overline\mu
=\bigl(\overline{A}_{\overline{R}+M},\overline{A}_{\overline{R}+M-1},\cdots,\overline{A}_{1},
A_1,A_2,\cdots,A_R|
\overline{B}_{\overline{R}},\overline{B}_{\overline{R}-1},\cdots,\overline{B}_{1},
B_1,B_2,\cdots,B_{R+M}\bigr),
\label{compsF}
\end{align}
as
\begin{align}
&\overline{A}_i=-\mu'_i+i-1-M,\quad
A_i=\lambda_i-i-M,\nonumber\\
&B_j=\lambda'_j-j+M,\quad
\overline{B}_j=-\mu_j+j-1+M,\nonumber\\
&R=\card\{i|A_i\ge 0\}=\card\{j|B_j\ge 0\}-M,\nonumber\\
&\overline{R}=\card\{i|\overline{A}_i\le -1\}-M=\card\{j|\overline{B}_j\le -1\}.
\label{composite}
\end{align}
We can apply exactly the same notation for both the $M\ge 0$ case and the $M\le 0$ case.
Note that, however, for the $M\le 0$ case, $R$ and $\overline{R}$ always satisfy $R\ge|M|$ and $\overline{R}\ge|M|$, so that the numbers of $B_j$ and $\overline A_i$ in the $M$-shifted Frobenius notation \eqref{compsF} are non-negative.
We can provide a pictorial interpretation for the arm lengths and the leg lengths (see figure \ref{compositeY} once again), by locating the other Young diagram $\mu$ in the composite Young diagram $\lambda\overline\mu$ in the opposite direction from $\lambda$.
Then the arm lengths and the leg lengths are naturally interpreted as the lengths to the rims of the composite Young diagram.
Note that we do not have to worry about the difference in the numbers of the arm lengths and the leg lengths in this definition.
The numbers of the shifted arm/leg lengths $A_i$, $B_j$ defined for $\lambda$ and those of $\overline{A}_i$, $\overline{B}_j$ defined for $\overline\mu$ originally have difference in $|M|$, though the total numbers are equal after taking the disjoint sum.
Note also that, the auxiliary arm or leg lengths for a single Young diagram need not be introduced separately, but appear naturally by setting one of the representations $\mu$ in \eqref{composite} to be the trivial one, no matter whether $M$ is positive or negative.
See figure \ref{RRbar} for a pictorial interpretation of $R$ and $\overline{R}$.

\subsection{ABJM matrix model}\label{abjm}

Now let us review the ABJM matrix model, especially focusing on the aspect of the integrable structure.
The ABJM theory \cite{ABJM,HLLLP2,ABJ} is the ${\cal N}=6$ superconformal Chern-Simons theory with gauge group U$(N_1)_k\times$U$(N_2)_{-k}$ (with $(k,-k)$ being the Chern-Simons levels) and two pairs of bifundamental matters.
This theory was identified to describe the worldvolume of $\min(N_1,N_2)$ M2-branes and $|N_2-N_1|$ fractional M2-branes on a target space ${\mathbb C}^4/{\mathbb Z}_k$. 
After utilizing the localization technique, the partition function and one-point functions of the half-BPS Wilson loop, which are originally defined with infinite-dimensional path integrals, reduce to finite-dimensional multiple integrals \cite{KWY}.
In the matrix model, the contribution from the gauge group and the bifundamental matters of the ABJM theory is combined perfectly into the invariant measure of the supergroup U$(N_1|N_2)$ and the half-BPS Wilson loop changes into the super Schur polynomial $s_\lambda(e^x|e^y)=s_\lambda(e^{x_1},\cdots,e^{x_{N_1}}|e^{y_1},\cdots,e^{y_{N_2}})$, the character of U$(N_1|N_2)$ \cite{MPtop,DT}.
Explicitly, the one-point function of the matrix model is given by
\begin{align}
&\langle s_\lambda\rangle_k(N_1,N_2)
=i^{-\frac{1}{2}(N_1^2-N_2^2)}
\int_{{\mathbb R}^{N_1+N_2}}\frac{D_k^{N_1}x}{N_1!}\frac{D_{-k}^{N_2}y}{N_2!}\nonumber\\
&\quad\times
\frac{\prod_{m<m'}^{N_1}(2\sinh\frac{x_m-x_{m'}}{2})^2
\prod_{n<n'}^{N_2}(2\sinh\frac{y_n-y_{n'}}{2})^2}
{\prod_{m=1}^{N_1}\prod_{n=1}^{N_2}(2\cosh\frac{x_m-y_n}{2})^2}
s_\lambda(e^x|e^y),
\end{align}
with the integrations accompanied with Fresnel factors as
\begin{align}
D_kx_m=\frac{dx_m}{2\pi}e^{\frac{ik}{4\pi}x_m^2},\quad
D_{-k}y_n=\frac{dy_n}{2\pi}e^{-\frac{ik}{4\pi}y_n^2}.
\end{align}
Note that, if we set the representation to be the trivial one, $\lambda=\bullet$, the one-point function reduces to the partition function $\langle s_\bullet\rangle_k(N_1,N_2)=\langle 1\rangle_k(N_1,N_2)$.

In the study of the matrix model, it is convenient to fix the rank difference $M=N_2-N_1$ and consider the grand canonical ensemble by regarding the rank as a particle number and introducing a dual fugacity $z$ as in
\begin{align}
\langle s_\lambda\rangle^\text{GC}_{k,M}(z)
=\sum_{N=\max(0,-M)}^\infty z^N\langle s_\lambda\rangle_k(N,N+M).
\end{align}
Then, it was found \cite{MatsumotoM} that the grand canonical one-point functions normalized by the grand canonical partition function without the rank difference
\begin{align}
S_\lambda^{M}=i^{-\frac{1}{2}M^2}\frac{\langle s_\lambda\rangle^\text{GC}_{k,M}(z)}{\langle 1\rangle^\text{GC}_{k,0}(z)},
\label{gcs}
\end{align}
satisfy the so-called shifted Giambelli relation.
Namely, if we prepare a suitable set of functions $\h_{(\widetilde A|\widetilde B)}$ labeled by two integers $(\widetilde A,\widetilde B)\in{\mathbb Z}\times{\mathbb Z}\setminus{\mathbb Z}_{<0}\times{\mathbb Z}_{<0}$, then, for any representation, $S_\lambda^{M}$ for $M\ge 0$ is given by
\begin{align}
S^{M\ge 0}_\lambda=\det\begin{pmatrix}
{\bigl(\h_{(\overline{A}_i|B_j)}\bigr)}
_{\begin{subarray}{l}M\ge i\ge 1\\1\le j\le M+R\end{subarray}}\\
{\bigl(\h_{(A_i|B_j)}\bigr)}
_{\begin{subarray}{l}1\le i\le R\\1\le j\le M+R\end{subarray}}
\end{pmatrix},
\label{sG+}
\end{align}
while $S_\lambda^{M}$ for $M\le 0$ is given by
\begin{align}
S^{M\le 0}_\lambda=\det\begin{pmatrix}
{\bigl(\h_{(A_i|\overline{B}_j)}\bigr)}
_{\begin{subarray}{l}1\le i\le R\\|M|\ge j\ge 1\end{subarray}}&
{\bigl(\h_{(A_i|B_j)}\bigr)}
_{\begin{subarray}{l}1\le i\le R\\1\le j\le M+R\end{subarray}}
\end{pmatrix},
\label{sG-}
\end{align}
where $A_i$, $B_j$ and $\overline A_i$, $\overline B_j$ are respectively the arm/leg lengths and the auxiliary lengths in the $M$-shifted Frobenius notation introduced in the previous subsection.
Here, for the order of the subscripts of submatrices, we have used the notation
\begin{align}
\bigl(M_{i,j}\bigr)
_{\begin{subarray}{l}I_1\gtreqless i\gtreqless I_N\\J_1\gtreqless j\gtreqless J_N\end{subarray}}
=\begin{pmatrix}
M_{I_1,J_1}&\cdots&M_{I_1,J_N}\\
\vdots&&\vdots\\
M_{I_N,J_1}&\cdots&M_{I_N,J_N}
\end{pmatrix}.
\end{align}
Namely, the subscripts are consecutive integers either in the increasing or decreasing order, though the leftmost/rightmost elements in the inequalities always indicate the first/last row or column.

There are some properties irrelevant to the explicit form of $\h_{(\widetilde A|\widetilde B)}$.
For example, with the expression of \eqref{sG+} and \eqref{sG-} we can prove the Jacobi-Trudi formula \cite{FM}
\begin{align}
S^M_{\lambda}/S^M_{\bullet}
=\det\begin{pmatrix}S^{M+j-1}_{[\lambda_i-i+j]}/S^{M+j-1}_{\bullet}\end{pmatrix},
\end{align}
without referring to the explicit components of $\h_{(\widetilde A|\widetilde B)}$.
Furthermore it was known in \cite{md} that with the Jacobi-Trudi formula we can prove the original Giambelli formula \cite{HHMO,MatsunoM}
\begin{align}
S^M_{\lambda}/S^M_{\bullet}=\det\begin{pmatrix}S^M_{(\alpha_i|\beta_j)}/S^M_{\bullet}\end{pmatrix},
\end{align}
if we denote the Young diagram by $\lambda=(\alpha_1,\cdots,\alpha_r|\beta_1,\cdots,\beta_r)$.
These formulas were originally found for the Schur polynomial, though it turns out that they appear naturally in integrable systems \cite{KOS,HL,AKLTZ,AZ}.
Hence, in this context, it is natural to ask what the integrable structure for the ABJM matrix model is.

Before proceeding to identifying the integrable structure, let us review two-point functions in the ABJM matrix model since the identification seems more natural in this framework.
Although the derivation of two-point functions with the localization technique is missing, in \cite{KM} the two-point function in the ABJM matrix model was defined as
\begin{align}
&\langle s_\lambda\overline s_\mu\rangle_k(N_1,N_2)
=i^{-\frac{1}{2}(N_1^2-N_2^2)}
\int_{{\mathbb R}^{N_1+N_2}}\frac{D_k^{N_1}x}{N_1!}\frac{D_{-k}^{N_2}y}{N_2!}\nonumber\\
&\quad\times
\frac{\prod_{m<m'}^{N_1}(2\sinh\frac{x_m-x_{m'}}{2})^2
\prod_{n<n'}^{N_2}(2\sinh\frac{y_n-y_{n'}}{2})^2}
{\prod_{m=1}^{N_1}\prod_{n=1}^{N_2}(2\cosh\frac{x_m-y_n}{2})^2}
s_\lambda(e^x|e^y)s_\mu(e^{-x}|e^{-y}).
\label{2ptdef}
\end{align}
Note that if we insert the super Schur polynomial of the other representation $\mu$ with the same arguments as the original one $\lambda$, $s_\mu(e^{x}|e^{y})$, the two insertions can be simply replaced by a single one using the Littlewood-Richardson expansion, $s_\lambda(e^x|e^y)s_\mu(e^x|e^y)=\sum_\nu N_{\lambda\mu}^\nu s_\nu(e^x|e^y)$.
This is why we have inverted the sign in the power of the arguments and inserted $s_\mu(e^{-x}|e^{-y})=s_\mu(e^{-x_1},\cdots,e^{-x_{N_1}}|e^{-y_1},\cdots,e^{-y_{N_2}})$.
Also, note that if we set $\mu=\bullet$ the two-point function reduces to the standard one-point function $\langle s_\lambda\overline s_\bullet\rangle_k(N_1,N_2)=\langle s_\lambda\rangle_k(N_1,N_2)$ (or the partition function if we further set $\lambda=\bullet$).
The definition of the two-point function seems to be a natural generalization from the one-point function.
Besides, as we review in the following, the two-point functions satisfy a generalization of the shifted Giambelli relations \eqref{sG+} and \eqref{sG-}.

After introducing the grand canonical ensemble
\begin{align}
\langle s_\lambda\overline s_\mu\rangle^\text{GC}_{k,M}(z)=\sum_{N=\max(0,-M)}^\infty z^N
\langle s_\lambda\overline s_\mu\rangle_k(N,N+M),
\end{align}
it was found that the normalized grand canonical two-point functions
\begin{align}
S_{\lambda\overline\mu}^M=i^{-\frac{1}{2}M^2}
\frac{\langle s_\lambda\overline s_{\mu}\rangle^\text{GC}_{k,M}(z)}{\langle 1\rangle^\text{GC}_{k,0}(z)},
\label{gcss}
\end{align}
satisfy a generalization of the shifted Giambelli relation for the one-point functions \cite{KM}.
Namely, given a suitable set of functions $\h_{(\widetilde A|\widetilde B)}$ labeled by two integers $(\widetilde A,\widetilde B)\in{\mathbb Z}\times{\mathbb Z}$, the (normalized grand canonical) two-point function $S_{\lambda\overline\mu}^{M}$ is given by
\begin{align}
S^{M}_{\lambda\overline\mu}=\det\begin{pmatrix}
{\bigl(\h_{(\overline{A}_i|\overline{B}_j)}\bigr)}
_{\begin{subarray}{l}\overline{R}+M\ge i\ge 1\\\overline{R}\ge j\ge 1\end{subarray}}&
{\bigl(\h_{(\overline{A}_i|B_j)}\bigr)}
_{\begin{subarray}{l}\overline{R}+M\ge i\ge 1\\1\le j\le M+R\end{subarray}}\\
{\bigl(\h_{(A_i|\overline{B}_j)}\bigr)}
_{\begin{subarray}{l}1\le i\le R\\\overline{R}\ge j\ge 1\end{subarray}}&
{\bigl(\h_{(A_i|B_j)}\bigr)}
_{\begin{subarray}{l}1\le i\le R\\1\le j\le M+R\end{subarray}}
\end{pmatrix}.
\label{2sG+}
\end{align}
Here, regardless of the sign of $M$, the subscripts of the components in \eqref{2sG+}, $A_i$, $B_j$ and $\overline A_i$, $\overline B_j$, are all of the arm lengths and the leg lengths in the $M$-shifted Frobenius notation of the composite Young diagram $\lambda\overline\mu$ \eqref{composite} and appear in the order of the blank arrows in figure \ref{compositeY}.
More compactly, with the $M$-shifted Frobenius notation \eqref{compsF} understood, the two-point function \eqref{gcss} is given as
\begin{align}
S^{M}_{\lambda\overline\mu}=\det\Bigl(\h_{(\widetilde A_i|\widetilde B_j)}\Bigr).
\label{2sG}
\end{align}
Since this is a natural generalization of the results for the one-point functions, this result implies that our definition of the two-point functions is a natural one.

Hence, to summarize, even though the derivation from the localization technique is missing, aside from the physical arguments given in \cite{KM}, we are convinced of the validity of the definition of the two-point functions from the following two viewpoints.
Namely, first, it is natural to define the two-point functions \eqref{2ptdef} by inserting the super Schur polynomials with the sign in the exponential functions inverted.
Secondly, the shifted Giambelli relation for the two-point functions \eqref{2sG} is a natural generalization of that for the one-point functions.

\subsection{Fermionic construction}

In this subsection, let us briefly summarize the results of the fermionic construction developed for studying the integrable hierarchy.
This was first formulated in \cite{DJKM,JM} and recent references \cite{AKLTZ,AZ} are also useful for us.
Following these references\footnote{We mainly follow the notation of \cite{JM,AZ}. Though the notation may not be the best one to see the correspondence to the matrix model, we try not to change so much from the original works.}, first we introduce generators of charged free fermions $\psi_n,\psi^*_n$ ($n\in\mathbb{Z}$) satisfying the Clifford algebra,
\begin{align}
\{\psi_m,\psi^*_n\}=\delta_{mn},\quad
\{\psi_m,\psi_n\}=\{\psi^*_m,\psi^*_n\}=0,\quad
\text{for}\quad m,n\in\mathbb{Z},
\end{align}
where $\psi_n$ has charge $+1$ and $\psi^*_n$ has charge $-1$.
The vacuum bra and ket states are determined by separating the free fermions into the creation operators and the annihilation operators.
Namely, the vacuum states $\langle n|$, $|n\rangle$ ($n$-vacuum states) are defined by
\begin{align}
\langle n|\psi^*_{m}=0,&\quad\text{for}\quad n>m,&
\psi_m|n\rangle=0,&\quad\text{for}\quad m<n,
\nonumber\\
\langle n|\psi_{m}=0,&\quad\text{for}\quad n\le m,&
\psi_m^*|n\rangle=0,&\quad\text{for}\quad m\ge n,
\end{align}
with $m,n\in\mathbb{Z}$, and are constructed by acting the creation operators to the $0$-vacuum states $\langle 0|$, $|0\rangle$ as
\begin{align}
\langle n|&=
\begin{cases}
\langle 0|\psi^*_0\psi^*_1\cdots\psi_{n-1}^*,&\text{for}\quad 0<n,\\
\langle 0|\psi_{-1}\psi_{-2}\cdots\psi_{n},&\text{for}\quad n<0,
\end{cases}&
|n\rangle&=\begin{cases}
\psi_{n-1}\cdots\psi_1\psi_0|0\rangle,&\text{for}\quad 0<n,\\
\psi^*_{n}\cdots\psi^*_{-2}\psi^*_{-1}|0\rangle,&\text{for}\quad n<0.
\end{cases}
\label{nexpand}
\end{align}
The excited states are given by acting the creation operators to the $n$-vacuum states where the action of the operators is encoded in the Young diagram as
\begin{align}
\langle\lambda,n|=\langle n|\psi^*_{n+\alpha_1}\cdots\psi^*_{n+\alpha_r}
\psi_{n-\beta_r-1}\cdots\psi_{n-\beta_1-1},\;
|\lambda,n\rangle=\psi^*_{n-\beta_1-1}\cdots\psi^*_{n-\beta_r-1}
\psi_{n+\alpha_r}\cdots\psi_{n+\alpha_1}|n\rangle,
\label{youngfermion}
\end{align}
through the Frobenius notation $\lambda=(\alpha_1,\cdots,\alpha_r|\beta_1,\cdots,\beta_r)$.
These states satisfy the orthogonal property
\begin{align}
\langle\lambda,m|\mu,n\rangle=\delta_{mn}\delta_{\lambda\mu}.
\end{align} 
Also, we can consider the multiplicative group of charge $0$
\begin{align}
\{G|{}^{\exists}G^{-1},G{\cal V}G^{-1}={\cal V},G{\cal V}^*G^{-1}={\cal V}^*\},
\label{Def_of_G}
\end{align}
mapping among ${\cal V}=\oplus_{n\in\mathbb{Z}}\mathbb{C}\psi_n$ and ${\cal V}^*=\oplus_{n\in\mathbb{Z}}\mathbb{C}\psi^*_n$.
From the definition we know that there is a matrix $(R_{mn})$ satisfying
\begin{align}
G\psi_nG^{-1}=(R^{-1}\psi)_n,\quad
G\psi^*_nG^{-1}=(\psi^*R)_n,
\label{Gadj}
\end{align}
where the expressions on the right-hand side are abbreviations
\begin{align}
(R^{-1}\psi)_n=\sum_{k\in\mathbb{Z}}{(R^{-1})}_{nk}\psi_k,\quad
(\psi^*R)_n=\sum_{k\in\mathbb{Z}}\psi^*_kR_{kn}.
\label{Rpsi}
\end{align}
We shall adopt these abbreviations when there is no confusion.
Also, the bilinear relation
\begin{align}
\sum_{k\in\mathbb{Z}}\psi_kG\otimes\psi^*_kG=\sum_{k\in\mathbb{Z}}G\psi_k\otimes G\psi^*_k,
\label{bc}
\end{align}
holds where a generalized version of the Wick's theorem in appendix \ref{wicktheorem} stems from.

We can construct the mKP tau function $\tau_n(t_{+})$ and the 2DTL tau function $\tau_n(t_{+},t_{-})$ with the fermionic construction
\begin{align}
\tau_n(t_{+})&=\langle n|e^{J_{+}(t_+)}G|n\rangle,\nonumber\\
\tau_n(t_{+},t_{-})&=\langle n|e^{J_{+}(t_+)}Ge^{-J_{-}(t_-)}|n\rangle,
\label{taufunc}
\end{align}
where the operators $J_\pm(t_\pm)$ are respectively defined as
\begin{align}
J_{\pm}(t_{\pm})=\sum_{k=1}^\infty t_{\pm k}J_{\pm k},\quad
J_k=\sum_{j\in\mathbb{Z}}\psi_j\psi^*_{j+k},
\label{Jpm}
\end{align}
with the arguments $t_\pm$ denoting a collection of $t_k$,
\begin{align}
t_{+}=\{t_{1},t_{2},\cdots\},\quad
t_{-}=\{t_{-1},t_{-2},\cdots\}.
\end{align}
By using these operators, we can also express the coefficients of the tau functions \eqref{taufunc}, when expanded by the Schur functions.
First we construct the function $\widetilde{s}_\lambda(t_\pm)$
\begin{align}
\widetilde{s}_\lambda(t_\pm)=\det\bigl(\widetilde h_{\lambda_i-i+j}(t_\pm)\bigr),
\label{sJT}
\end{align}
from the functions $\widetilde h_k(t_\pm)$ defined by
\begin{align}
\sum_{k=0}^\infty\widetilde h_k(t_\pm)z^k=e^{\xi(t_\pm,z)},\quad
\xi(t_\pm,z)=\sum_{k=1}^\infty t_{\pm k}z^k.
\end{align}
Since the function $\widetilde h_k(t_+)$ is nothing but the complete symmetric function $h_k(x_1,x_2,\cdots)$ under the substitution
\begin{align}
t_k=\frac{1}{k}(x^k_1+x^k_2+\cdots),
\end{align}
the function $\widetilde{s}_\lambda(t_+)$ defined from the Jacobi-Trudi formula \eqref{sJT} are equivalent to the Schur function $s_\lambda(x_{1},x_{2},\cdots)$.
Hence, with a slight abuse of terminology, we often refer to $\widetilde{s}_\lambda(t_\pm)$ itself also as the Schur function.
Using the operators $J_\pm(t_\pm)$ introduced in \eqref{Jpm}, we can express the Schur functions as
\begin{align}
{(-1)}^{b(\lambda)}\widetilde{s}_\lambda(t_{+})&=\langle n|e^{J_{+}(t_+)}|\lambda,n\rangle,
\nonumber\\
{(-1)}^{b(\lambda)}\widetilde{s}_\lambda(t_{-})&=\langle\lambda,n|e^{J_{-}(t_-)}|n\rangle,
\label{schurf}
\end{align} 
with $b(\lambda)=\sum_{j=1}^{r}(\beta_j+1)$.
For the derivation, here we have used the relations between $J_{\pm}(t_\pm)$ and $\psi_{n},\psi^*_{n}$,
\begin{align}
e^{J_{+}(t_+)}\psi_ne^{-J_{+}(t_+)}&=\sum_{k=0}^\infty\psi_{n-k}\widetilde h_k(t_+),&
e^{J_{+}(t_+)}\psi^*_ne^{-J_{+}(t_+)}&=\sum_{k=0}^\infty\psi^*_{n+k}\widetilde h_k(-t_+),\nonumber\\
e^{J_{-}(t_-)}\psi_ne^{-J_{-}(t_-)}&=\sum_{k=0}^\infty\psi_{n+k}\widetilde h_k(t_-),&
e^{J_{-}(t_-)}\psi^*_ne^{-J_{-}(t_-)}&=\sum_{k=0}^\infty\psi^*_{n-k}\widetilde h_k(-t_-).
\end{align}
which are derived from
\begin{align}
e^{J_{+}(t_+)}\psi(z)e^{-J_+(t_+)}&=e^{\xi(t_+,z)}\psi(z),&
e^{J_{+}(t_+)}\psi^*(z)e^{-J_+(t_+)}&=e^{-\xi(t_+,z)}\psi^*(z),\nonumber\\
e^{J_{-}(t_-)}\psi(z)e^{-J_-(t_-)}&=e^{\xi(t_-,z^{-1})}\psi(z),&
e^{J_{-}(t_-)}\psi^*(z)e^{-J_-(t_-)}&=e^{-\xi(t_-,z^{-1})}\psi^*(z),
\end{align}
with $\psi(z)$, $\psi^*(z)$ being the generating functions of $\psi_n$, $\psi^*_n$,
\begin{align}
\psi(z)&=\sum_{n\in\mathbb{Z}}\psi_nz^n,&
\psi^*(z)&=\sum_{n\in\mathbb{Z}}\psi^*_nz^{-n}.
\end{align}
Then, the tau functions \eqref{taufunc} can be expanded using the basis of the Schur functions \eqref{schurf} as
\begin{align}
\tau_n(t_{+})
&=\sum_{\lambda}c_{\lambda}(n)\widetilde{s}_\lambda(t_{+}),
\nonumber\\
\tau_n(t_{+},t_{-})
&=\sum_{\lambda,\mu}c_{\lambda\overline\mu}(n)
\widetilde{s}_\lambda(t_{+})\widetilde{s}_\mu(-t_{-}).
\label{tau-function}
\end{align}
where the coefficients $c_{\lambda}(n)$ and $c_{\lambda\overline\mu}(n)$ are
\begin{align}
c_{\lambda}(n)&={(-1)}^{b(\lambda)}\langle\lambda,n|G|n\rangle,
\nonumber\\
c_{\lambda\overline\mu}(n)&={(-1)}^{b(\lambda)+b(\mu)}\langle\lambda,n|G|\mu,n\rangle.
\label{cfermion}
\end{align}

At this point we already see many similarities between the stories in the ABJM matrix model in section \ref{abjm} and the fermionic construction in the current section.
First, if we set one of the Young diagram to be trivial $\mu=\bullet$, we obtain the reduction from the 2DTL hierarchy to the mKP hierarchy, $c_{\lambda\overline\bullet}(n)=c_{\lambda}(n)$.
Secondly, the coefficients of the mKP hierarchy, $c_{\lambda}(n)$, satisfy the Pl\"ucker relations, such as the Giambelli formula and the Jacobi-Trudi formula \cite{AKLTZ,AZ}.
We shall see the correspondence more explicitly in the next section.

\section{ABJM/2DTL correspondence}

In the previous section, we have reviewed the two-point functions in the ABJM matrix model and the fermionic construction for the 2DTL hierarchy.
In this section, we claim that they have the same integrable structure.

\subsection{Correspondence}\label{correspond}

Our proposal of the identification is that the (normalized grand canonical) two-point function in the ABJM matrix model \eqref{gcss} corresponds to the coefficient of the 2DTL tau function \eqref{cfermion} as
\begin{align}
S_{\lambda\overline\mu}^{M}={(-1)}^{\frac{1}{2}n(n+1)}
\frac{c_{\lambda\overline\mu}(n)}{c_{\bullet}(0)}\bigg|_{n=-M}.
\label{identify}
\end{align}
Namely from the viewpoint of the two-point function, the rank difference $M$ in $S_{\lambda\overline\mu}^M$ is identified to be the charge of the vacuum by
\begin{align}
n=-M,
\end{align}
and the two representations in the two-point function are identified with the excitations from the vacua respectively for the bra state and the ket state.
By reduction it is clear that, for $\mu =\bullet$, the one-point function in the ABJM matrix model \eqref{gcs} corresponds to the coefficient of the mKP tau function \eqref{cfermion} as
\begin{align}
S^{M}_{\lambda}={(-1)}^{\frac{1}{2}n(n+1)}\frac{c_{\lambda}(n)}{c_{\bullet}(0)}\bigg|_{n=-M}.
\end{align}

Our strategy to see the correspondence explicitly is as follows.
One of the main difficulties is that the coefficient of the tau function \eqref{cfermion} along with \eqref{youngfermion} is originally expressed in the Frobenius notation, though the two-point function is given in the $M$-shifted Frobenius notation \eqref{2sG}.
Hence, in the next subsection, we first rewrite the expression in terms of the $M$-shifted Frobenius notation.
Then in section \ref{compABJM}, we choose special cases of the rank difference $M$ and the Young diagrams $\lambda$, $\mu$ in the two-point function of the ABJM matrix model, so that only one component of $\h_{(\widetilde A|\widetilde B)}$ in \eqref{2sG} appears.
Since the two-point function is fixed uniquely from $\h_{(\widetilde A|\widetilde B)}$ with the shifted Giambelli relation, we prove that the same shifted Giambelli relation for the fermionic construction in section \ref{sGfermi}.
The proof relies on a generalization of the Wick's theorem, which we present in section \ref{gWick} and prove in appendix \ref{wicktheorem}.

Before closing this subsection, let us elaborate on the generating function \eqref{generating} in the introduction.
Under the identification \eqref{identify} with the original expression for $S_{\lambda\overline\mu}^M$ \eqref{gcss}, we find that the relation
\begin{align}
c_{\lambda\overline{\mu}}(n)\simeq(-1)^{\frac{1}{2}n(n+1)}i^{-\frac{1}{2}{n}^2}
\langle s_\lambda\overline s_\mu\rangle^\text{GC}_{k,-n},
\end{align}
holds up to a numerical factor independent of the composite Young diagram $\lambda\overline{\mu}$ and the rank difference $M=-n$.
Due to this reason, the generating function of the two-point functions in the ABJM matrix model \eqref{generating}
\begin{align}
F_n(t_+,t_-)=\sum_{\lambda,\mu}\Bigl[(-1)^{\frac{1}{2}n(n+1)}i^{-\frac{1}{2}{n}^2}
\langle s_\lambda\overline s_\mu\rangle^\text{GC}_{k,-n}\Bigr]
\widetilde{s}_\lambda(t_+)\widetilde{s}_\mu(-t_{-}),
\end{align}
can be identified as the 2DTL tau function \eqref{tau-function} and therefore satisfies various equations in the 2DTL hierarchy such as \eqref{2dtleq} (see, for example, \cite{AZ} for a derivation of \eqref{2dtleq}).

\subsection{Tau function in shifted Frobenius notation}\label{taushift}

In identifying the two-point function in the fermionic construction of the 2DTL hierarchy in \eqref{identify}, we have used the coefficient \eqref{cfermion} of the tau function expanded by the Schur functions,
\begin{align}
c_{\lambda\overline{\mu}}(n)&={(-1)}^{b(\lambda)+b(\mu)}\langle\lambda,n|G|\mu,n\rangle,
\label{bb}
\end{align}
with
\begin{align}
b(\lambda)&=\sum_{j=1}^{r}(\beta_j+1),&
\langle\lambda,n|&=\langle n|\psi^*_{n+\alpha_1}\cdots\psi^*_{n+\alpha_r}
\psi_{n-\beta_r-1}\cdots\psi_{n-\beta_1-1},\nonumber\\
b(\mu)&=\sum_{i=1}^{\overline{r}}\overline{\alpha}_i,&
|\mu,n\rangle&=\psi^*_{n+\overline{\alpha}_1}\cdots\psi^*_{n+\overline{\alpha}_{\overline{r}}}
\psi_{n-\overline{\beta}_{\overline{r}}-1}\cdots\psi_{n-\overline{\beta}_1-1}|n\rangle.
\label{cdefs}
\end{align}
Here we regard $\mu$ as a part of the composite Young diagram $\lambda\overline\mu$ and adopt the Frobenius notation for $\lambda\overline\mu$ \eqref{Fcomp} to describe \eqref{youngfermion}.
Since the two-point function in the ABJM matrix model is given by the shifted Frobenius notation, our first task in the identification is to rewrite the expression for $c_{\lambda\overline{\mu}}(n)$ in terms of the shifted Frobenius notation,
\begin{align}
&c_{\lambda\overline{\mu}}(n)\Big|_{n=-M}
=(-1)^{\sum_{j=1}^{M+R}(B_j+1)+\sum_{i=1}^{\overline{R}+M}\overline{A}_i}\nonumber\\
&\times
\langle 0|\psi^*_{A_1}\cdots\psi^*_{A_R}
\psi_{-B_{M+R}-1}\cdots\psi_{-B_1-1}G
\psi^*_{\overline{A}_1}\cdots\psi^*_{\overline{A}_{\overline{R}+M}}
\psi_{-\overline{B}_{\overline{R}}-1}\cdots\psi_{-\overline{B}_1-1}|0\rangle.
\label{csF+}
\end{align}
Note that this rewriting is valid regardless of the sign of $M$.

\begin{figure}[!t]
\centering\includegraphics[scale=0.6,angle=90]{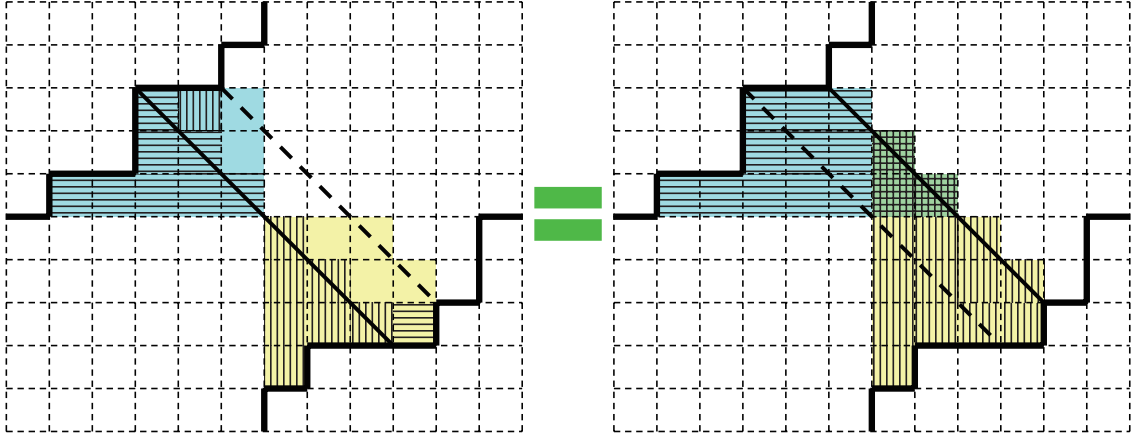}
\caption{A pictorial interpretation of the computation of the signs in \eqref{signcomp}.
Each term can be interpreted as the area in the composite Young diagram.
On the left-hand side of \eqref{signcomp},
$\sum_{j=1}^r(\beta_j+1)$, $\sum_{i=1}^{\overline r}\overline\alpha_i$,
$MR$, $\sum_{i=R+1}^{r}\alpha_i$,
$-M\overline R$ and $\sum_{i=\overline R+1}^{\overline r}(\overline\beta_i+1)$
are respectively regarded as the numbers of yellow boxes with vertical strips, cyan boxes with horizontal strips, yellow boxes without strips, yellow boxes with horizontal strips, cyan boxes without strips and cyan boxes with vertical strips.
On the right-hand side, $\sum_{i=1}^{M+R}(B_i+1)$ and $\sum_{i=1}^{\overline{R}+M}\overline{A}_i$ are respectively the total number of both yellow boxes and green boxes and that of both cyan boxes and green boxes.
Here the contributions from the green boxes cancel each other.
}
\label{signY}
\end{figure}

Let us explain the rewriting for $M=-n\ge 0$.
After substituting the expression of the $n$-vacuum state $\langle n|$ in \eqref{nexpand}, the bra state $\langle\lambda,n|$ is given by
\begin{align}
\langle\lambda,n|\Big|_{n=-M}
&=\langle 0|\psi_{-1}\cdots\psi_{-M}\psi^*_{-M+\alpha_1}\cdots\psi^*_{-M+\alpha_r}
\psi_{-M-\beta_r-1}\cdots\psi_{-M-\beta_1-1}.
\end{align}
Then, we change the order of the fermions so that $\psi^*$'s are located in the left and $\psi$'s are in the right as in the original expression in \eqref{cdefs}.
In bringing the sequence $\psi_{-1}\cdots\psi_{-M}$ to the right, some of $\psi^*_{-M+\alpha_i}$ are annihilated and some are not.
After anti-commuting the $R$ fermions $\psi^*_{-M+\alpha_i}$ ($1\le i\le R$) with the sequence $\psi_{-1}\cdots\psi_{-M}$ obtaining the sign factor $(-1)^{MR}$, the remaining $(r-R)$ fermions $\psi^*_{-M+\alpha_i}$ ($R+1\le i\le r$) are all annihilated by the sequence $\psi_{-1}\cdots\psi_{-M}$.
For the annihilation to work directly, we need to anti-commute $\psi^*_{-M+\alpha_i}$ with the product $\psi_{-M+\alpha_i-1}\cdots\psi_{-M}$ in the sequence $\psi_{-1}\cdots\psi_{-M}$ giving the sign factor $(-1)^{\alpha_i}$, starting from $i=R+1$ until $i=r$.
Finally, we combine the remaining $M-(r-R)$ fermions out of the sequence $\psi_{-1}\cdots\psi_{-M}$ with the existing ones $\psi_{-M-\beta_r-1}\cdots\psi_{-M-\beta_1-1}$ by
\begin{align}
&\{-B_{M+R}-1,\cdots,-B_1-1\}\nonumber\\
&=\bigl(\{-1,\cdots,-M\}\backslash\{-M+\alpha_{R+1},\cdots,-M+\alpha_r\}\bigr)
\sqcup\{-M-\beta_r-1,\cdots,-M-\beta_1-1\},
\end{align}
to obtain
\begin{align}
\langle\lambda,n|\Big|_{n=-M}={(-1)}^{MR+\sum_{i=R+1}^{r}\alpha_i}
\langle 0|\psi^*_{A_1}\cdots\psi^*_{A_R}\psi_{-B_{M+R}-1}\cdots\psi_{-B_1-1}.
\label{braexcite}
\end{align}
The same rewriting applies to the ket state $|\mu,n\rangle$
\begin{align}
|\mu,n\rangle\Big|_{n=-M}=(-1)^{-M\overline R+\sum_{j=\overline R+1}^{\overline r}(\overline\beta_j+1)}
\psi^*_{\overline A_1}\cdots\psi^*_{\overline A_{\overline R+M}}
\psi_{-\overline B_{\overline R}-1}\cdots\psi_{-\overline B_1-1}|0\rangle.
\label{ketexcite}
\end{align}
Finally we collect the sign factors from \eqref{bb}, \eqref{braexcite} and \eqref{ketexcite} to find
\begin{align}
(-1)^{\sum_{j=1}^r(\beta_j+1)+\sum_{i=1}^{\overline r}\overline\alpha_i
+MR+\sum_{i=R+1}^{r}\alpha_i
-M\overline R+\sum_{i=\overline R+1}^{\overline r}(\overline\beta_i+1)}
=(-1)^{\sum_{i=1}^{M+R}(B_i+1)+\sum_{i=1}^{\overline{R}+M}\overline{A}_i},
\label{signcomp}
\end{align}
where the computation is easily understood by counting the box numbers in the composite Young diagram (see figure \ref{signY} for an explanation).
The same rewriting works for $M=-n\le 0$ as well.

\subsection{Component in ABJM matrix model}\label{compABJM}

\begin{figure}[!t]
\centering\includegraphics[scale=0.4,angle=90]{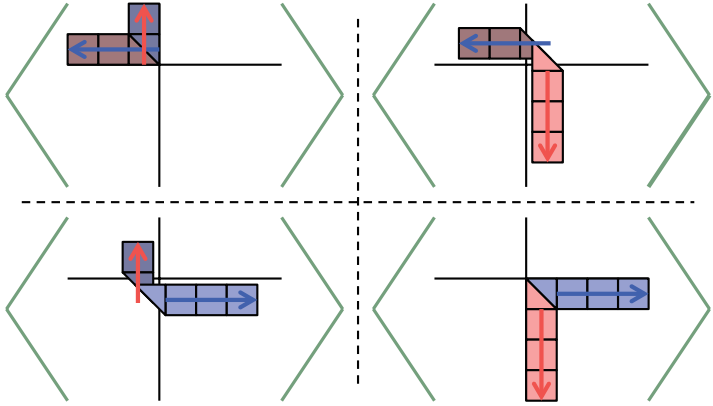}
\caption{The four types of correlation functions in the ABJM matrix model where the determinant of the shifted Giambelli relation \eqref{2sG} contains only one component $\h_{(\widetilde A|\widetilde B)}$.
The upper-left, upper-right, lower-left and lower-right figures denote respectively $S^0_{\bullet,\overline{(1|2)}}=\h_{(-3|-2)}$, $S^1_{(0|2),\overline{(0|1)}}=\h_{(-3|3)}$, $S^{-1}_{(2|0),\overline{(0|0)}}=\h_{(3|-2)}$ and $S^0_{(3|3),\overline{\bullet}}=\h_{(3|3)}$.
The blue arrow and the red arrow denote respectively the arm length and the leg length in the shifted Frobenius notation, while the blue box and the red box are related to the arm length and the leg length of the two Young diagrams with the light color and the dark color used respectively for $\lambda$ and $\mu$ in the composite Young diagram $\lambda\overline\mu=(\lambda,\mu)$.
Though the expression in \eqref{shifthook} seems cumbersome, the correlation functions in the ABJM matrix model with only one component of $\h_{(\widetilde A|\widetilde B)}$ are simply the hook types in terms of the composite Young diagram.
}
\label{sgh}
\end{figure}

After we propose the identification in section \ref{correspond} and rewrite the fermionic construction in the shifted Frobenius notation in section \ref{taushift}, we can provide an explicit expression for the component $\h_{(\widetilde A|\widetilde B)}$ which is used to express the two-point function in the ABJM matrix model with the shifted Giambelli relation \eqref{2sG}.

For this purpose, let us apply the identification \eqref{identify} to the cases where the determinant of the shifted Giambelli relation contains only one component.
Namely for $\h_{(\overline{A}|\overline{B})}$, $\h_{(\overline{A}|B)}$, $\h_{(A|\overline{B})}$ and $\h_{(A|B)}$ we respectively choose
\begin{align}
(\lambda,\mu,M)&=\bigl(\bullet,({-\overline B-1}|{-\overline A-1}),0\bigr),&
(\lambda,\mu,M)&=\bigl((0|B-1),(0|{-\overline A-2}),1\bigr),\nonumber\\
(\lambda,\mu,M)&=\bigl((A-1|0),({-\overline B-2}|0),-1\bigr),&
(\lambda,\mu,M)&=\bigl((A|B),\bullet,0\bigr),
\label{shifthook}
\end{align}
in the Frobenius notation.
Although the choice of the pairs of the Young diagrams looks random at first sight, these are simply the hook types in terms of the composite Young diagram,
\begin{align}
S^0_{\bullet,\overline{({-\overline B-1}|{-\overline A-1})}}&=\h_{(\overline A|\overline B)},&
S^{1}_{(0|B-1),\overline{(0|{-\overline A-2})}}&=\h_{(\overline{A}|B)},\nonumber\\
S^{-1}_{(A-1|0),\overline{({-\overline B-2}|0)}}&=\h_{(A|\overline{B})},&
S^0_{(A|B),\overline{\bullet}}&=\h_{(A|B)},
\label{h2pt}
\end{align}
(see figure \ref{sgh}; appendix \ref{compositeG} and figure \ref{shiftG} are also helpful).
Then, from the identification \eqref{identify}, the components $\h_{(\widetilde A|\widetilde B)}$ for the shifted Giambelli relation are given by
\begin{align}
\h_{(\overline{A}|\overline{B})}
&={(-1)}^{\overline{A}}\frac{\langle G\psi^{*}_{\overline{A}}\psi_{-\overline{B}-1}\rangle}
{\langle G\rangle},&
\h_{(\overline{A}|{B})}
&={(-1)}^{B+1+\overline{A}}\frac{\langle\psi_{-B-1}G\psi^{*}_{\overline{A}}\rangle}
{\langle G\rangle},\nonumber\\
\h_{({A}|\overline{B})}
&=-\frac{\langle\psi^{*}_{A}G\psi_{-\overline{B}-1}\rangle}{\langle G\rangle},&
\h_{({A}|{B})}
&={(-1)}^{B+1}\frac{\langle\psi^{*}_{A}\psi_{-B-1}G\rangle}{\langle G\rangle},
\label{hcomponent}
\end{align}
where, for convenience, we have introduced a simplified notation of the vacuum expectation values $\langle\cdots\rangle$ as
\begin{align}
\langle\cdots\rangle:=\langle\bullet,0|\cdots|\bullet,0\rangle=\langle 0|\cdots|0\rangle.
\end{align}

Now note that, through the shifted Giambelli relation \eqref{2sG}, the two-point function in the ABJM matrix model is fixed unambiguously by $\h_{(\widetilde A|\widetilde B)}$.
Therefore our identification of $\h_{(\widetilde A|\widetilde B)}$ from certain two-point functions in this section already fixes all the two-point functions in principle.
The question of identifying the two-point functions in the ABJM matrix model in the fermionic construction, now changes into the question whether the identification \eqref{identify} is consistent with the shifted Giambelli relation \eqref{2sG}.
In section \ref{sGfermi}, we see explicitly that the fermionic construction satisfies the shifted Giambelli relation.
However, before that, we need to prepare a generalization of the Wick's theorem.

\subsection{Generalized Wick's theorem}\label{gWick}

In the proof of the shifted Giambelli relation for the coefficients of the tau function, we need a generalization of the Wick's theorem, which we present in this subsection.

It is well-known that the Wick's theorem holds for the free theories.
Especially for the free fermions a version of it was proved in \cite{AKLTZ}.
Namely, let $v_j=\sum_kv_{jk}\psi_{k}$ be a linear combination of $\psi_k$ and $w_i^*=\sum_k\psi_k^*w_{ki}^*$ be a linear combination of $\psi_k^*$.
Then, the version of it
\begin{align}
\frac{\langle n|v_r\cdots v_1w_1^*\cdots w_r^* G|n\rangle}{\langle n|G|n\rangle}
=\det\biggl(\frac{\langle n|v_jw_i^* G|n\rangle}{\langle n|G|n\rangle}\biggr)
_{\begin{subarray}{c}1\le i\le r\\1\le j\le r\end{subarray}},
\label{AKLTZWick}
\end{align}
holds where $G$ is a group element defined in \eqref{Def_of_G}.
Here we prove a generalization of it, where the relative order of $v_j$ and $w^*_i$ can be arbitrary while among $\{v_j\}_{j=1}^{r}$ and among $\{w^*_i\}_{i=1}^{r}$ themselves the orders are always $v_{r},\cdots,v_1$ and $w_1^*,\cdots,w_{r}^*$ respectively.
This generalization is important for our application in proving the shifted Giambelli relation for the fermionic construction.

\begin{figure}[!t]
\centering\includegraphics[scale=0.5,angle=-90]{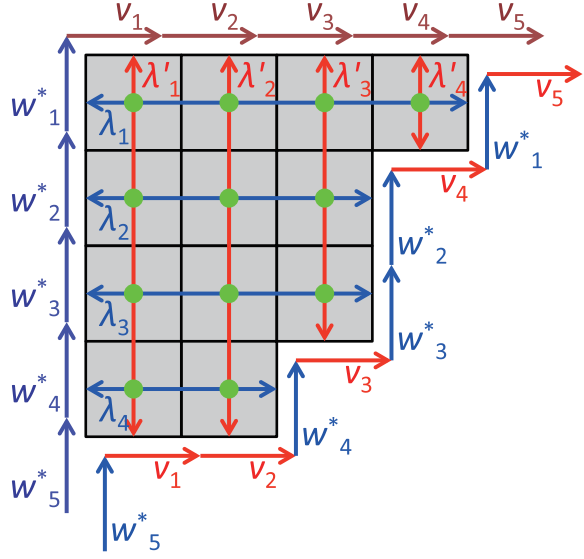}
\caption{Example of the product
$v_5w_1^*v_4w_2^*w_3^*v_3w_4^*v_2v_1w_5^*$.
We denote $w_i^*$ and $v_j$ by the upward arrows and the rightward arrows and connect them successively.
$\lambda_i$ and $\lambda'_j$ denote respectively the deviations from the product in the reference order $v_5v_4v_3v_2v_1w_1^*w_2^*w_3^*w_4^*w_5^*$
and therefore are the horizontal and vertical lengths of the displacements, $[\lambda_1,\lambda_2,\lambda_3,\lambda_4,\lambda_5]=[4,3,3,2,0]$ and $[\lambda'_1,\lambda'_2,\lambda'_3,\lambda'_4,\lambda'_5]'=[4,4,3,1,0]'$.
The sign of the permutation can be found by counting the intersections of arrows (green dots) showing the displacements, which is in one-to-one correspondence with the boxes where the intersections are located.
}
\label{order}
\end{figure}

We first prepare a notation to explain the generalization.
For the product of $m+n$ operators $\{v_j\}_{j=1}^n\sqcup\{w_i^*\}_{i=1}^m$, it is easiest to express the relative order of $v_j$ and $w^*_i$ by associating it with the Young diagram as follows (see figure \ref{order}).
Namely, we read the product of operators from the right to the left and associate the operator $w_i^*$ to be an arrow going upward and $v_j$ to be an arrow going rightward.
Then, the product $v_nv_{n-1}\cdots v_2v_1w_1^*w_2^*\cdots w_{m-1}^*w_m^*$ such as that in \eqref{AKLTZWick} corresponds to the Young diagram in the trivial representation and we refer to it as the reference order.
For a product in a general order, the polygonal lines consisting of the up-going arrows and the right-going arrows form the rim of the Young diagram.
With this association to the Young diagram, $w_i^*$ and $v_j$ are respectively deviated from those in the reference order by $\lambda_i$ and $\lambda'_j$ and located at the $(m+1-i+\lambda_i)$-th position and at the $(m+j-\lambda'_j)$-th position counting from the right.
We further denote the order of $w^*_i$ and $v_j$ by $\lambda$ and introduce a notation $[v_n\cdots v_1;w_1^*\cdots w_m^*]_\lambda$ to be the product of the $m+n$ operators in this order multiplied by the relative sign of the permutation from the reference order $v_n\cdots v_1w_1^*\cdots w_m^*$.
The sign of the permutation is determined by the number of transpositions from the product in the reference order, which is obtained by counting the intersections (green dots in figure \ref{order}) of the arrows connecting two corresponding operators.
Since each intersection corresponds to the box where the intersection is located, the number of transpositions is nothing but the total box number of the Young diagram, therefore the sign can be denoted by $(-1)^{\sum_{i=1}^m\lambda_i}=(-1)^{\sum_{j=1}^n\lambda'_j}$.
For example, for the product in the order $v_5w_1^*v_4w_2^*w_3^*v_3w_4^*v_2v_1w_5^*$, we find
\begin{align}
[v_5v_4v_3v_2v_1;w^*_1w^*_2w^*_3w^*_4w^*_5]_{[4,3,3,2]}
=(-1)^{4+3+3+2}v_5w_1^*v_4w_2^*w_3^*v_3w_4^*v_2v_1w_5^*.
\end{align}
Then, we propose a generalization of the Wick's theorem as
\begin{align}
\frac{\langle n|[v_r\cdots v_1;w_1^*\cdots w_r^*]_\lambda G|n\rangle}{\langle n|G|n\rangle}
=\det\biggl(\frac{\langle n|[v_j;w_i^*]_\lambda G|n\rangle}{\langle n|G|n\rangle}\biggr)
_{\begin{subarray}{c}1\le i\le r\\1\le j\le r\end{subarray}}.
\label{generalWick}
\end{align}
Here, for $[v_j;w_i^*]_\lambda$ on the right-hand side, where the product consisting only of a subset of operators in the original order, we understand the order $\lambda$ in $[\cdots]_\lambda$ as a restriction of $\lambda$ to the subset,
\begin{align}
[v_j;w_i^*]_\lambda=\begin{cases}v_jw_i^*,
&\text{for}\quad
[v_r\cdots v_1;w_1^*\cdots w_r^*]_\lambda=\pm\cdots v_j\cdots w_i^*\cdots,\\
-w_i^*v_j,
&\text{for}\quad
[v_r\cdots v_1;w_1^*\cdots w_r^*]_\lambda=\pm\cdots w_i^*\cdots v_j\cdots.\end{cases}
\label{vwtwo}
\end{align}
The proof of \eqref{generalWick} is given in appendix \ref{wicktheorem}.

\subsection{Shifted Giambelli relation in fermionic construction}\label{sGfermi}

In this subsection we prove that the coefficients \eqref{csF+} of the tau function in the fermionic construction satisfy the same shifted Giambelli relation as in the two-point functions \eqref{2sG}.
Then, we can find the consistency of the identification \eqref{identify}, by comparing the shifted Giambelli relation for both the two-point functions and the fermionic construction.
For the proof we utilize the Wick's theorem \eqref{generalWick} given in the previous subsection.

To apply the Wick's theorem, we first collect the fermions in the coefficient of the tau function \eqref{csF+} into one side of the group element $G$ using \eqref{Gadj},
\begin{align}
&\frac{c_{\lambda\overline\mu}(n)}{c_{\bullet}(0)}\bigg|_{n=-M}
=(-1)^{\sum_{j=1}^{M+R}(B_j+1)+\sum_{i=1}^{\overline{R}+M}\overline{A}_i}
\langle 0|\psi^*_{A_1}\cdots\psi^*_{A_R}
\psi_{-B_{M+R}-1}\cdots\psi_{-B_1-1}\nonumber\\
&\quad\times
(\psi^*R)_{\overline{A}_1}\cdots(\psi^*R)_{\overline{A}_{M+\overline{R}}}
(R^{-1}\psi)_{-\overline{B}_{\overline{R}}-1}\cdots(R^{-1}\psi)_{-\overline{B}_1-1}G|0\rangle
/\langle G\rangle.
\end{align}
Here we have used the simplified notation introduced in \eqref{Rpsi}.
Then we can apply the Wick's theorem \eqref{generalWick}.
Since the fermions in the vacuum expectation value are deviated from the reference order given in the previous subsection, we need to include the sign factor $(-1)^{R(M+R)+R\overline{R}+\overline{R}(\overline{R}+M)}$ in applying the Wick's theorem.
Then, the Wick's theorem states that the vacuum expectation value is given by a determinant
\begin{align}
&\frac{c_{\lambda\overline\mu}(n)}{c_{\bullet}(0)}\bigg|_{n=-M}
=(-1)^{\sum_{j=1}^{M+R}(B_j+1)+\sum_{i=1}^{\overline{R}+M}\overline{A}_i
+R(M+R)+R\overline{R}+\overline{R}(\overline{R}+M)}\nonumber\\
&\quad\times\det\begin{pmatrix}
\bigl(-\langle\psi^*_{A_i}(R^{-1}\psi)_{-\overline B_j-1}G\rangle/\langle G\rangle\bigr)
_{\begin{subarray}{l}1\le i\le R\\1\le j\le\overline R\end{subarray}}&
\bigl(-\langle\psi^*_{A_i}\psi_{-B_j-1}G\rangle/\langle G\rangle\bigr)
_{\begin{subarray}{l}1\le i\le R\\1\le j\le M+R\end{subarray}}\\
\bigl(-\langle(\psi^*R)_{\overline A_i}(R^{-1}\psi)_{-\overline B_j-1}G\rangle/\langle G\rangle\bigr)
_{\begin{subarray}{l}1\le i\le\overline R+M\\1\le j\le\overline R\end{subarray}}&
\bigl(\langle\psi_{-B_j-1}(\psi^*R)_{\overline A_i}G\rangle/\langle G\rangle\bigr)
_{\begin{subarray}{l}1\le i\le\overline R+M\\1\le j\le M+R\end{subarray}}
\end{pmatrix}.
\end{align}
Then, we follow several steps of changes.
We change the order of operators so that the fermions are located on the correct side of the group element $G$; we change the rows and the columns so that they are aligned in the same order as \eqref{2sG}; we then distribute the sign factor $(-1)^{\sum_{j=1}^{M+R}(B_j+1)+\sum_{i=1}^{\overline{R}+M}\overline{A}_i}$ to each column and row to reproduce the components of the determinant \eqref{hcomponent}.
Finally, we find
\begin{align}
&(-1)^{\frac{1}{2}n(n+1)}\frac{c_{\lambda\overline\mu}(n)}{c_{\bullet}(0)}\bigg|_{n=-M}
\nonumber\\
&=\det\begin{pmatrix}
\bigl((-1)^{\overline A_i}\langle G\psi^*_{\overline A_i}\psi_{-\overline B_j-1}\rangle/\langle G\rangle\bigr)
_{\begin{subarray}{l}\overline R+M\ge i\ge 1\\\overline R\ge j\ge 1\end{subarray}}&
\bigl((-1)^{B_j+1+\overline A_i}\langle\psi_{-B_j-1}G\psi^*_{\overline A_i}\rangle/\langle G\rangle\bigr)
_{\begin{subarray}{l}\overline R+M\ge i\ge 1\\1\le j\le M+R\end{subarray}}\\
\bigl(-\langle\psi^*_{A_i}G\psi_{-\overline B_j-1}\rangle/\langle G\rangle\bigr)
_{\begin{subarray}{l}1\le i\le R\\\overline R\ge j\ge 1\end{subarray}}&
\bigl((-1)^{B_j+1}\langle\psi^*_{A_i}\psi_{-B_j-1}G\rangle/\langle G\rangle\bigr)
_{\begin{subarray}{l}1\le i\le R\\1\le j\le M+R\end{subarray}}
\end{pmatrix}.
\label{sGfermion}
\end{align}
By further substituting the expression for the components $\h_{(\widetilde A|\widetilde B)}$ \eqref{hcomponent} (obtained by applying the identification \eqref{identify} to the hook types) to the right-hand side, we obtain exactly the shifted Giambelli relation for the two-point function $S^{M}_{\lambda\overline\mu}$ \eqref{2sG}.
Since this argument holds for any composite Young diagram $\lambda\overline\mu$ and any rank difference $M=-n$, we finally prove that the identification \eqref{identify} of the two-point functions in the fermionic construction is consistent.

To summarize, the two-point functions in the ABJM matrix model share the integrable structure with the 2DTL hierarchy.
Our proof of the consistency of the identification \eqref{identify} relies, however, on the fact that there exists a group element $G$ in \eqref{Def_of_G} satisfying the expression \eqref{hcomponent}.
To see this, in the next subsection, we present an explicit construction of the group element $G$ in terms of the two-point functions $\h_{(\widetilde A|\widetilde B)}$ \eqref{h2pt} using \eqref{hcomponent}.

\subsection{Construction of group element}

In section \ref{compABJM} and especially \eqref{hcomponent}, we have learned that the components $\h_{(\widetilde A|\widetilde B)}$ in the shifted Giambelli relation \eqref{2sG} are given by the vacuum expectation values of the group element $G$, which is characterized by the matrix $R=(R_{mn})$ via \eqref{Gadj}.
In this subsection, we determine the group element $G$ (or the matrix $R$) inversely from the two-point functions in the ABJM matrix model $\h_{(\widetilde A|\widetilde B)}$ \eqref{h2pt}.

We first rewrite the group element $G$ into a more useful expression to compute \eqref{hcomponent}.
Since the bilinear sum of fermions $\sum_{m,n\in\mathbb{Z}}{\mathcal{B}}_{mn}\psi^*_m\psi_n$ of charge $0$ generates the algebra gl($\infty$), we can express the group element $G$ of GL($\infty$) as
\begin{align}
G=e^{\sum_{m,n\in\mathbb{Z}}{\mathcal{B}}_{mn}\psi^*_m\psi_n},
\end{align}
where the matrix $\mathcal{B}=(\mathcal{B}_{mn})$ is related to the matrix $R$ by
\begin{align}
R=e^{\mathcal{B}}.
\end{align}
Furthermore, since the expression of the fermionic construction in \eqref{hcomponent} is given with the $0$-vacuum state, it is convenient to express the group element $G$ with the normal-ordering,
\begin{align}
G=\langle G\rangle\;{}_\bullet^\bullet e^{\sum_{m,n\in\mathbb{Z}}{\mathcal{A}}_{mn}\psi^*_m\psi_n}{}_\bullet^\bullet,
\label{Gexp}
\end{align}
where ${}_\bullet^\bullet\cdots{}_\bullet^\bullet$ denotes the normal-ordering for the 0-vacuum state $|0\rangle$ and the matrix $\mathcal{A}=({\mathcal{A}}_{mn})$ is determined by the matrix ${\mathcal B}$ or $R$ as \cite{AKLTZ,AZ} 
\begin{align}
\mathcal{A}=(R-I){\bigl(I+P_{+}(R-I)\bigr)}^{-1},
\label{AR}
\end{align}
with $I$ and $P_+$ being respectively the identity and the projection
\begin{align}
I_{mn}=\delta_{mn},\quad
{(P_{+})}_{mn}=\begin{cases}
\delta_{mn},&\quad\text{for}\quad m,n\ge 0,\\
0,&\quad\text{otherwise}.\end{cases}
\end{align}
Then, using \eqref{Gexp}, the components $\h_{(\widetilde A|\widetilde B)}$, expressed in terms of the vacuum expectation values of $G$ in \eqref{hcomponent}, can be computed explicitly as
\begin{align}
\h_{(\overline{A}|\overline{B})}
&={(-1)}^{\overline{A}}{\bigl({(R^{-1})}_{-}{\mathcal{A}}_{+}R\bigr)}_{-\overline{B}-1,\overline{A}},
&\h_{(\overline{A}|{B})}
&={(-1)}^{B+1+\overline{A}}{\bigl(R-{\mathcal{A}}_{+}R\bigr)}_{-B-1,\overline{A}},\nonumber\\
\h_{({A}|\overline{B})}
&={-\bigl((R^{-1})+{(R^{-1})}_{-}\mathcal{A}\bigr)}_{-\overline{B}-1,A},
&\h_{({A}|{B})}
&={(-1)}^{B+1}{\mathcal{A}}_{-B-1,A},
\label{Rtoh}
\end{align}
where the products between matrices are given by
\begin{align}
\bigl({\mathcal{A}}_+R\bigr)_{mn}=\sum_{k\in\mathbb{Z}_{\ge 0}}{\mathcal{A}}_{mk}R_{kn},\quad
\bigl({(R^{-1})}_{-}\mathcal{A}\bigr)_{mn}=\sum_{k\in\mathbb{Z}_{<0}}{(R^{-1})}_{mk}\mathcal{A}_{kn}.
\end{align}
Note that, since ${\mathcal A}$ is given in terms of $R$ \eqref{AR}, the whole expression of $\h_{(\widetilde A|\widetilde B)}$ in \eqref{Rtoh} is given only by the matrix $R$.
In the way we obtain a complete relation between the two-point functions (characterized by $\h_{(\widetilde A|\widetilde B)}$) and the fermionic construction (characterized by the matrix $R$).
Here the derivation of \eqref{Rtoh} can be performed as follows.
We first bring the group element $G$ to the rightmost using \eqref{Gadj} so that it acts directly on the $0$-vacuum ket state.
Then, in the vacuum expectation values with only two fermions, the exponential part of the group element $G$ \eqref{Gexp} only contributes as
$G=\langle G\rangle(1+\sum_{m<0,n\ge 0}{\cal A}_{mn}\psi_m^*\psi_n)$.

To provide an explicit relation, in the following we introduce a notation for an infinite matrix $M$ by dividing it into four parts,
\begin{align}
M=\begin{pmatrix}
M_{--}&M_{-+}\\
M_{+-}&M_{++}
\end{pmatrix}
=\begin{pmatrix}
\bigl(M_{mn}\bigr)_{\begin{subarray}{l}m\le -1\\n\le -1\end{subarray}}
&\bigl(M_{mn}\bigr)_{\begin{subarray}{l}m\le -1\\0\le n\end{subarray}}\\
\bigl(M_{mn}\bigr)_{\begin{subarray}{l}0\le m\\n\le -1\end{subarray}}
&\bigl(M_{mn}\bigr)_{\begin{subarray}{l}0\le m\\0\le n\end{subarray}}
\end{pmatrix}.
\end{align}
Also, we define a modified matrix $H=(H_{mn})$ $(m,n\in\mathbb{Z})$ by the components $\h_{({\widetilde{A}}|{\widetilde{B}})}$ in the shifted Giambelli relation as
\begin{align}
H_{mn}=(-1)^{\frac{1}{2}(m-|m|)+\frac{1}{2}(n-|n|)}\h_{(n|-m-1)}.
\end{align}
Using these notations, the relations \eqref{Rtoh} are expressed as
\begin{align}
H=\begin{pmatrix}
H_{--}&H_{-+}\\
H_{+-}&H_{++}
\end{pmatrix}
&=\begin{pmatrix}
\bigl(R-\mathcal{A}_+R\bigr)_{--}
&\mathcal{A}_{-+}\\
\bigl((R^{-1})_-\mathcal{A}_+R\bigr)_{+-}
&-\bigl((R^{-1})+(R^{-1})_-\mathcal{A}\bigr)_{++}
\end{pmatrix}.
\end{align}
If we divide the matrix $R$ into four parts in the same way,
\begin{align}
R=\begin{pmatrix}
R_{--}&R_{-+}\\
R_{+-}&R_{++}
\end{pmatrix},
\end{align}
we can express each part of the matrix $H$ more explicitly as
\begin{align}
H_{--}&=R_{--}-R_{-+}(R_{++})^{-1}R_{+-},
&H_{-+}&=R_{-+}(R_{++})^{-1},\nonumber\\
H_{+-}&=-(\Delta_{++})^{-1}R_{+-}(R_{--})^{-1}R_{-+}(R_{++})^{-1}R_{+-},
&H_{++}&=-(R_{++})^{-1},
\label{HinR}
\end{align}
with $\Delta_{++}=R_{++}-R_{+-}(R_{--})^{-1}R_{-+}$.
In the derivation, the following expressions for the matrices $R^{-1}$ and $\mathcal{A}$ are useful,
\begin{align}
R^{-1}
&=\begin{pmatrix}
(R_{--})^{-1}+(R_{--})^{-1}R_{-+}(\Delta_{++})^{-1}R_{+-}(R_{--})^{-1}&
-(R_{--})^{-1}R_{-+}(\Delta_{++})^{-1}\\
-(\Delta_{++})^{-1}R_{+-}(R_{--})^{-1}&
(\Delta_{++})^{-1}\\
\end{pmatrix},\nonumber\\
\mathcal{A}
&=\begin{pmatrix}
-I_{--}+R_{--}-R_{-+}(R_{++})^{-1}R_{+-}&R_{-+}(R_{++})^{-1}\\
(R_{++})^{-1}R_{+-}&I_{++}-(R_{++})^{-1}\\
\end{pmatrix}.
\end{align}

By solving \eqref{HinR} for $R$ inversely, the matrix $R$ is given in terms of the two-point functions in the ABJM matrix model as
\begin{align}
R_{--}&=H_{--}\Bigl(1+\sqrt{(H_{--})^{-1}H_{-+}(H_{++})^{-1}H_{+-}}\Bigr),
&R_{-+}&=-H_{-+}(H_{++})^{-1},\nonumber\\
R_{+-}&=(H_{-+})^{-1}H_{--}\sqrt{(H_{--})^{-1}H_{-+}(H_{++})^{-1}H_{+-}},
&R_{++}&=-(H_{++})^{-1}.
\end{align}
Here, for the square root of the infinite matrix, we need to choose a suitable branch.

\section{Conclusion}

We have proved that the two-point functions in the ABJM matrix model share exactly the same integrable structure as the two-dimensional Toda lattice hierarchy.
Since the two-point function in the ABJM matrix model is given by the shifted Giambelli relation, after identifying each component in the fermionic construction, we prove that the fermionic construction satisfies the same shifted Giambelli relation and the identification is consistent.
Putting in other words, we have identified the generating function of the two-point functions as the solution (tau-function) of the two-dimensional Toda lattice hierarchy.
It was known that gauge theories with large number of supersymmetries often enjoy the integrable structure, such as the anomalous dimension of the trace operators in the four-dimensional ${\cal N}=4$ super Yang-Mills theory \cite{YM}.
We are able to add one more example to the list.

As we have mentioned in the introduction, the derivation of the two-point function from the localization technique is missing.
Hence, aside from the physical arguments given in \cite{KM}, the proposal on the two-point function in the ABJM matrix model is also based on the aesthetic viewpoint of the shifted Giambelli relation.
In this work we have clarified further the structure of the two-point functions as the two-dimensional Toda lattice hierarchy and its reduction to the one-point functions as the modified KP hierarchy.

After our identification of the two-point functions in the ABJM matrix model in the fermionic construction, this implies infinitely many relations for the system.
Especially, in appendix \ref{compositeG}, we propose and prove a novel relation, a generalization of the Giambelli formula for the two-point functions.

Let us raise several directions to pursue further.

First, the identification of the ABJM matrix model as the integrable hierarchy implies infinitely many non-linear differential equations.
We would like to understand the physical interpretation of these equations.

Secondly, we have identified the integrable structure of the ABJM matrix model to be the two-dimensional Toda lattice hierarchy using the fermionic construction.
It was also known that the grand canonical partition function of the ABJM matrix model (and its generalizations) can be rewritten into that of a fermionic system \cite{MP,MatsumotoM,MN1,KM}.
Both of these two fermionic systems are related to the same determinant expression of the ABJM matrix model.
The determinant may be reminiscent of the Slavnov determinant appearing in the structure constant of the four-dimensional ${\cal N}=4$ super Yang-Mills theory \cite{EGSV,F}, which is also connected to the integrable hierarchy \cite{FS,T}.
It would be interesting to clarify the relations.

Thirdly, after the identification, the Giambelli formula \cite{HHMO,MatsunoM} and the Jacobi-Trudi formula \cite{FM} for the one-point functions are proved in a more systematic way \cite{AKLTZ}.
Besides these formulas implying the integrability, there are, however, many other relations satisfied by the ABJM matrix model which are not explained from the shifted Giambelli relation.
For example, the explicit expression for the components $\h_{(\widetilde A|\widetilde B)}$ is not independent and satisfies the conjugate relations \cite{KM}.
Other examples such as the open-closed duality \cite{HaOk,KiyoshigeM} are also not explained directly from the integrable structure.
It is interesting to understand these relations from the integrable viewpoint.

Fourthly, though we have mainly focused on the open string side so far, the story on the closed string side is interesting as well.
The large $z$ expansion of the grand canonical partition function was studied in a series of works \cite{DMP1,HKPT,DMP2,FHM,MP,KEK,HMO2,CM,HMO3,HMMO,MatsumotoM,HO} and it was found to be expressed by the free energy of the closed topological string theory on local ${\mathbb P}^1\times{\mathbb P}^1$.
Interestingly enough, in the analysis \cite{MP,HMO1,PY,HMO2,CM} an integrable structure \cite{TW1,TW2} originating from the polymer matrix model \cite{Z} was utilized.
This relation is generalized to many other geometries \cite{GHM1,KaMa,MZ,KMZ,WZH,GKMR,CGrM,SWH,CGuM} and many other superconformal Chern-Simons matrix models \cite{MN3,HHO,MNN,MNY,KMN}.
Besides, interesting relations such as the Wronskian relation \cite{GHM2} (with the chiral projections interpreted as the orientifold \cite{H,MS2,MN5}) or the $q$-Painlev\'e equation \cite{BGT} implying an integrable structure \cite{KNY}, were also proposed.
We hope that our identification of the integrable structure is helpful to understand the relation to topological strings or $q$-Painlev\'e equations.

Finally, the integrable structure in matrix models is not new.
For example, the integrable hierarchy was known for a series of standard matrix models \cite{ZJ,ADKMV,DV}.
We would like to understand the relation to these matrix models.

\appendix

\section{Proof of generalized Wick's theorem}\label{wicktheorem}

In this appendix, let us prove the generalized Wick's theorem \eqref{generalWick},
\begin{align}
\frac{\langle n|[v_r\cdots v_1;w_1^*\cdots w_r^*]_\lambda G|n\rangle}{\langle n|G|n\rangle}
=\det\biggl(\frac{\langle n|[v_j;w_i^*]_\lambda G|n\rangle}{\langle n|G|n\rangle}\biggr)
_{\begin{subarray}{c}1\le i\le r\\1\le j\le r\end{subarray}},
\end{align}
by induction in a similar way as \eqref{AKLTZWick} in \cite{AKLTZ} utilizing the bilinear identity \eqref{bc}
\begin{align}
\sum_k\langle U|\psi_kG|V\rangle\langle U'|\psi_k^* G|V'\rangle
=\sum_k\langle U|G\psi_k|V\rangle\langle U'|G\psi_k^*|V'\rangle,
\label{bbc}
\end{align}
and the commutation relation
\begin{align}
v_j\psi_k^* =v_{jk}-\psi_k^* v_j,\quad
w_i^*\psi_k=w_{ki}^*-\psi_kw_i^*.
\label{cr}
\end{align}

\begin{figure}[!t]
\centering\includegraphics[scale=0.5,angle=-90]{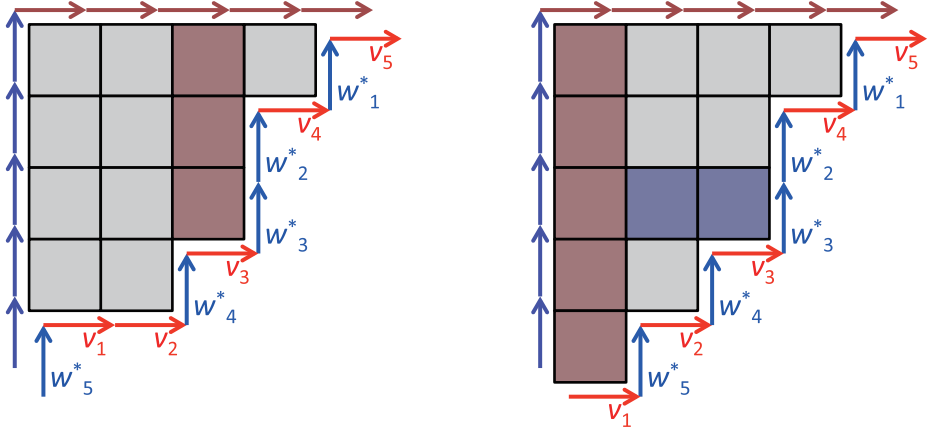}
\caption{A pictorial method to understand the extra signs in removing elements from the product $[v_r\cdots v_1;w_1^*\cdots w_r^*]_\lambda$.
(Left)
In the removal of $w^*_5$ no extra signs appear, $[v_5\cdots v_1;w_1^*\cdots w_5^*]_\lambda=[v_5\cdots v_1;w_1^*\cdots w_4^*]_\lambda w^*_5$, while the removal of $v_{j=3}$ causes an extra sign, $(-1)^{\lambda'_{j=3}}=(-1)^3$, as can be seen explicitly from $[v_5\cdots v_1;w_1^*\cdots w_5^*]_\lambda=(-1)^{12}v_5w_1^*v_4w_2^*w_3^*v_3w_4^*v_2v_1w_5^*$, $[v_5\cdots v_1;w_1^*\cdots w_4^*]_\lambda=(-1)^{12}v_5w_1^*v_4w_2^*w_3^*v_3w_4^*v_2v_1$ and $[v_5v_4\widecheck v_3v_2v_1;w_1^*w_2^*w_3^*w_4^*]_\lambda=(-1)^{9}v_5w_1^*v_4w_2^*w_3^*w_4^*v_2v_1$.
(Right)
The removal of $v_1$ causes an extra sign $(-1)^r=(-1)^5$ in $[v_5\cdots v_1;w_1^*\cdots w_5^*]_\lambda=(-1)^5[v_5\cdots v_2;w_1^*\cdots w_5^*]_\lambda v_1$.
After that, if we remove $w^*_{i=3}$ furthermore, an extra sign $(-1)^{\lambda_{i=3}-1}=(-1)^2$ appears.
}
\label{vw}
\end{figure}

On one hand, let us first consider the case when the rightmost element in the product is $w_r^*$,
\begin{align}
[v_r\cdots v_1;w_1^*\cdots w_r^*]_\lambda
=[v_r\cdots v_1;w_1^*\cdots w_{r-1}^*]_\lambda w^*_r,
\label{removew}
\end{align}
where the sign of the product $[v_r\cdots v_1;w_1^*\cdots w_{r-1}^*]_\lambda$ consisting only of a subset of operators in $\lambda$ is understood as in \eqref{vwtwo}.
By applying \eqref{bbc} with
\begin{align}
\langle U|=\langle n|w_r^*,\quad
\langle U'|=\langle n|[v_r\cdots v_1;w_1^*\cdots w_{r-1}^*]_\lambda,\quad
|V\rangle=|V'\rangle=|n\rangle,
\end{align}
we find
\begin{align}
\sum_k\langle n|w_r^*\psi_k G|n\rangle
\langle n|[v_r\cdots v_1;w_1^*\cdots w_{r-1}^*]_\lambda\psi_k^*G|n\rangle=0.
\end{align}
Note that the right-hand side vanishes since either $\psi_k^*$ or $\psi_k$ annihilates a fixed state $|n\rangle$ regardless of the choice of the state $|n\rangle$.
Now let us bring the oscillators $\psi_k^*$ and $\psi_k$ to the leftmost using \eqref{cr}.
For the first factor $\langle n|w_r^*\psi_k G|n\rangle$, we find
\begin{align}
\langle n|G|n\rangle
\langle n|[v_r\cdots v_1;w_1^*\cdots w_{r-1}^*]_\lambda w_r^*G|n\rangle
=\sum_k
\langle n|\psi_k w_r^* G|n\rangle
\langle n|[v_r\cdots v_1;w_1^*\cdots w_{r-1}^*]_\lambda\psi_k^*G|n\rangle.
\end{align}
If we bring $\psi_k^*$ to the leftmost in the second factor $\langle n|[v_r\cdots v_1;w_1^*\cdots w_{r-1}^*]_\lambda\psi_k^*G|n\rangle$ as well, the commutation relation \eqref{cr} gives a non-trivial contribution with $v_j$ which is located at the $(r-1+j-\lambda'_j)$-th position from the right.
When we describe the result with $v_j$ eliminated in the original notation $[v_r\cdots\widecheck v_j\cdots v_1;w^*_1\cdots w^*_{r-1}]_\lambda$, we have an extra sign $(-1)^{\lambda'_j}$ due to the missing in the total box number (see figure \ref{vw} for a pictorial method to understand the extra signs).
Thus the sign is $(-1)^{(r-1+j-\lambda'_j-1)+\lambda'_j}=(-1)^{r-j}$ and the final result is
\begin{align}
&\langle n|G|n\rangle
\langle n|[v_r\cdots v_1;w_1^*\cdots w_r^*]_\lambda G|n\rangle
\nonumber\\
&=\sum_j(-1)^{r-j}
\langle n|[v_j;w^*_r]_\lambda G|n\rangle
\langle n|[v_r\cdots\widecheck v_j\cdots v_1;w^*_1\cdots w^*_{r-1}]_\lambda G|n\rangle,
\label{wlast}
\end{align}
where we have changed $v_jw^*_r$ on the right-hand side into $v_jw^*_r=[v_j;w^*_r]_\lambda$ since $w_r^*$ is originally located in the rightmost.

On the other hand, when the rightmost element is $v_1$ in the product,
\begin{align}
[v_r\cdots v_1;w_1^*\cdots w_r^*]_\lambda 
=(-1)^{r}[v_r\cdots v_2;w_1^*\cdots w_r^*]_\lambda v_1,
\end{align}
(note that an extra sign appears by eliminating the first row in the Young diagram, see figure \ref{vw}), we can apply \eqref{bbc} with
\begin{align}
\langle U|=\langle n|[v_r\cdots v_2;w_1^*\cdots w_r^*]_\lambda,\quad
\langle U'|=\langle n|v_1,\quad
|V\rangle=|V'\rangle=|n\rangle.
\end{align}
As previously, by bringing $\psi_k$ to the leftmost in the second factor using \eqref{cr}, this implies
\begin{align}
\langle n|[v_r\cdots v_2;w_1^*\cdots w_r^*]_\lambda v_1G|n\rangle
\langle n|G|n\rangle
=\sum_k
\langle n|[v_r\cdots v_2;w_1^*\cdots w_r^*]_\lambda\psi_kG|n\rangle
\langle n|\psi_k^* v_1G|n\rangle.
\end{align}
Furthermore, when we bring $\psi_k$ to the leftmost in the first factor, it has the non-trivial commutation relation with $w^*_i$ at the $(r-i+\lambda_i)$-th position from the right.
If we express the result with $w^*_i$ removed in the original notation $[v_r\cdots v_2;w^*_1\cdots\widecheck w^*_i\cdots w^*_r]_\lambda$, we have an extra sign $(-1)^{\lambda_i-1}$ (see figure \ref{vw}).
Totally the sign is $(-1)^{r+(r-i+\lambda_i-1)+(\lambda_i-1)}=(-1)^{i}$.
Finally we find
\begin{align}
&\langle n|[v_r\cdots v_1;w_1^*\cdots w_r^*]_\lambda G|n\rangle
\langle n|G|n\rangle
\nonumber\\
&=\sum_i(-1)^{i-1}
\langle n|[v_r\cdots v_2;w^*_1\cdots\widecheck w^*_i\cdots w^*_r]_\lambda G|n\rangle
\langle n|[v_1;w_i^*]_\lambda G|n\rangle,
\label{vlast}
\end{align}
where we have used $w_i^*v_1=-[v_1;w_i^*]_\lambda$.
By combining \eqref{wlast} and \eqref{vlast} we can prove \eqref{generalWick} with the Laplace expansion by induction.

\section{Giambelli formula for composite Young diagram}\label{compositeG}

In this appendix, we propose a different type of the Giambelli formula.
Typically the Giambelli formula is considered for single Young diagrams (or the one-point functions in the ABJM matrix model) and the proof is obtained from the Wick's theorem.
After we prove the generalization of the Wick's theorem in appendix \ref{wicktheorem}, we can apply it to composite Young diagrams (the two-point functions).

\begin{figure}[!t]
\centering\includegraphics[scale=0.6,angle=-90]{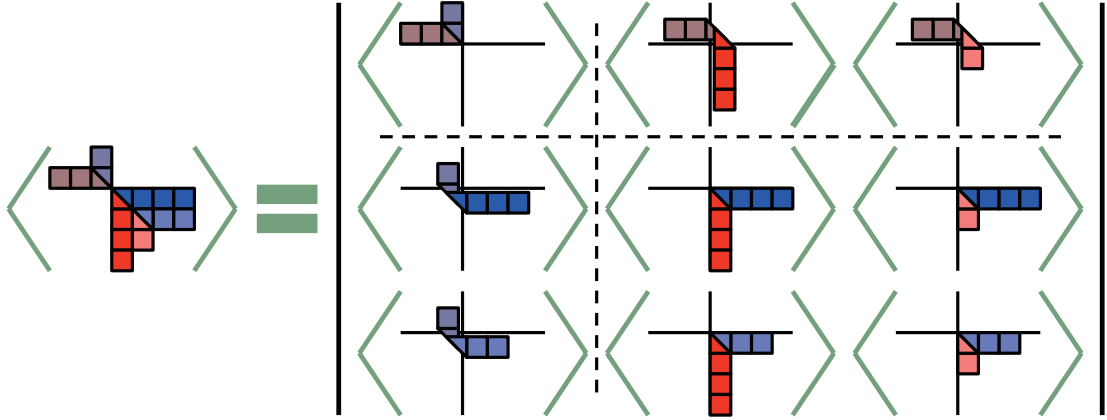}
\caption{Giambelli formula for the two-point functions in the ABJM matrix model.
}
\label{shiftG}
\end{figure}

Instead of change the $n$-vacuum state $\langle n|,|n\rangle$ into the $0$-vacuum state $\langle 0|,|0\rangle$ in proving the shifted Giambelli relation in section \ref{sGfermi}, for the proof of the Giambelli formula, we can directly apply the Wick's theorem \eqref{generalWick} for the composite Young diagram constructed on the $n$-vacuum state.
Then, as in section \ref{sGfermi}, we easily find
\begin{align}
&\frac{\langle\lambda,n|G|\mu,n\rangle}{\langle n|G|n\rangle}
=\det\begin{pmatrix}
{\displaystyle\biggl(
\frac{\langle n|G\psi^*_{n+\overline{\alpha}_i}\psi_{n-\overline{\beta}_j-1}|n\rangle}
{\langle n|G|n\rangle}
\biggr)}
_{\begin{subarray}{l}\overline{r}\ge i\ge 1\\\overline{r}\ge j\ge 1\end{subarray}}
&{\displaystyle\biggl(
\frac{\langle n|\psi_{n-\beta_j-1}G\psi^*_{n+\overline{\alpha}_i}|n\rangle}
{\langle n|G|n\rangle}
\biggr)}
_{\begin{subarray}{l}\overline{r}\ge i\ge 1\\1\le j\le r\end{subarray}}
\\
{\displaystyle\biggl(
-\frac{\langle n|\psi^*_{n+\alpha_i}G\psi_{n-\overline{\beta}_j-1}|n\rangle}
{\langle n|G|n\rangle}
\biggr)}
_{\begin{subarray}{l}1\le i\le r\\\overline{r}\ge j\ge 1\end{subarray}}
&{\displaystyle\biggl(
\frac{\langle n|\psi^*_{n+\alpha_i}\psi_{n-\beta_j-1}G|n\rangle}
{\langle n|G|n\rangle}
\biggr)}
_{\begin{subarray}{l}1\le i\le r\\1\le j\le r\end{subarray}}
\end{pmatrix},
\end{align}
which implies, by distributing the signs $(-1)^{b(\lambda)+b(\mu)}$,
\begin{align}
&\frac{c_{\lambda\overline{\mu}}(n)}{c_\bullet(n)}
=\det\begin{pmatrix}
{\displaystyle\biggl(\frac{c_{(\overline{\alpha}_i|\overline{\beta}_j)}(n)}{c_\bullet(n)}\biggr)}
_{\begin{subarray}{l}\overline{r}\ge i\ge 1\\\overline{r}\ge j\ge 1\end{subarray}}
&{\displaystyle\biggl(\frac{c_{(\overline{\alpha}_i+1,0|-1,\beta_j-1)}(n-1)}{c_\bullet(n)}\biggr)}
_{\begin{subarray}{l}\overline{r}\ge i\ge 1\\1\le j\le r\end{subarray}}
\\
{\displaystyle\biggl(-\frac{c_{(-1,\alpha_i-1|\overline{\beta}_j+1,0)}(n+1)}{c_\bullet(n)}\biggr)}
_{\begin{subarray}{l}1\le i\le r\\\overline{r}\ge j\ge 1\end{subarray}}
&{\displaystyle\biggl(\frac{c_{(\alpha_i|\beta_j)}(n)}{c_\bullet(n)}\biggr)}
_{\begin{subarray}{l}1\le i\le r\\1\le j\le r\end{subarray}}
\end{pmatrix}.
\end{align}
From the ABJM/2DTL correspondence \eqref{identify}, we can rewrite this formula in terms of the two-point functions in the ABJM matrix model as
\begin{align}
&\frac{S_{\lambda\overline{\mu}}^{M}}{S_\bullet^{M}}
=\det\begin{pmatrix}
{\displaystyle\biggl(\frac{S_{(\overline{\alpha}_i|\overline{\beta}_j)}^{M}}{S_\bullet^{M}}\biggr)}
_{\begin{subarray}{l}\overline{r}\ge i\ge 1\\\overline{r}\ge j\ge 1\end{subarray}}
&{\displaystyle\biggl(\frac{S_{(\overline{\alpha}_i+1,0|-1,\beta_j-1)}^{M+1}}{S_\bullet^{M}}\biggr)}
_{\begin{subarray}{l}\overline{r}\ge i\ge 1\\1\le j\le r\end{subarray}}
\\
{\displaystyle\biggl(\frac{S_{(-1,\alpha_i-1|\overline{\beta}_j+1,0)}^{M-1}}{S_\bullet^{M}}\biggr)}
_{\begin{subarray}{l}1\le i\le r\\\overline{r}\ge j\ge 1\end{subarray}}
&{\displaystyle\biggl(\frac{S_{(\alpha_i|\beta_j)}^{M}}{S_\bullet^{M}}\biggr)}
_{\begin{subarray}{l}1\le i\le r\\1\le j\le r\end{subarray}}
\end{pmatrix},
\end{align}
(see figure \ref{shiftG}) where the composite Young diagrams are represented in the Frobenius notation \eqref{Fcomp} and the charge $n$ of the $n$-vacuum $|n\rangle$ in the fermionic construction is identified with the rank difference $M$ by $n=-M$.
The reason why we call this formula the Giambelli formula for the two-point functions is that this formula is expressed in terms of the hook types of the composite Young diagram and that this reduces to the usual Giambelli formula for the one-point functions if we set $\mu=\bullet$.

\section*{Acknowledgements}

We are grateful to Omar Foda, Naotaka Kubo, Marcos Marino, Kazunobu Maruyoshi, Satsuki Matsuno, Masatoshi Noumi, Makoto Sakaguchi, Jun-ichi Sakamoto, Yasuhiko Yamada and Konstantin Zarembo for valuable discussions and comments.
The work of S.M.\ is supported by JSPS Grant-in-Aid for Scientific Research (C) \#26400245.
S.M.\ would like to thank Yukawa Institute for Theoretical Physics at Kyoto University for warm hospitality.


\begin{thebibliography}{10}
\bibitem{ZK}
N.~J.~Zabusky and M.~D.~Kruskal,
``Interaction of `Solitons' in a Collisionless Plasma and the Recurrence of Initial States,''
Phys.\ Rev.\ Lett.\  {\bf 15}, 240 (1965).
%doi:10.1103/PhysRevLett.15.240
%%CITATION = doi:10.1103/PhysRevLett.15.240;%%
%
\bibitem{S}
M.~Sato,
``Soliton equations as dynamical systems on infinite dimensional Grassmann manifolds,  (random systems and dynamical systems).''
RIMS Kokyuroku {\bf 439}, 30-46 (1981).
%
\bibitem{DJKM}
E.~Date, M.~Jimbo, M.~Kashiwara, T.~Miwa,
``Transformation groups for soliton equations,''
``Nonlinear integrable systems - classical theory and quantum theory,''
39-119, World Scientific Publishing, (1983).
%
\bibitem{MJD}
T.~Miwa, M.~Jimbo, E.~Date,
``Solitons: Differential equations, symmetries and infinite dimensional algebras,''
Vol.~135, Cambridge University Press, (2000).
%
\bibitem{JM}
M.~Jimbo, T.~Miwa, 
``Solitons and infinite dimensional Lie algebras,''
Publications of the Research Institute for Mathematical Sciences, {\bf 19} (3), 943-1001 (1983).
%
\bibitem{ABJM}
O.~Aharony, O.~Bergman, D.~L.~Jafferis and J.~Maldacena,
``N=6 superconformal Chern-Simons-matter theories, M2-branes and their gravity duals,''
JHEP {\bf 0810}, 091 (2008)
%doi:10.1088/1126-6708/2008/10/091
[arXiv:0806.1218 [hep-th]].
%%CITATION = doi:10.1088/1126-6708/2008/10/091;%%
%
\bibitem{HLLLP2}
K.~Hosomichi, K.~M.~Lee, S.~Lee, S.~Lee and J.~Park,
``N=5,6 Superconformal Chern-Simons Theories and M2-branes on Orbifolds,''
JHEP {\bf 0809}, 002 (2008)
%doi:10.1088/1126-6708/2008/09/002
[arXiv:0806.4977 [hep-th]].
%%CITATION = doi:10.1088/1126-6708/2008/09/002;%%
%
\bibitem{ABJ}
O.~Aharony, O.~Bergman and D.~L.~Jafferis,
``Fractional M2-branes,''
JHEP {\bf 0811}, 043 (2008)
%doi:10.1088/1126-6708/2008/11/043
[arXiv:0807.4924 [hep-th]].
%%CITATION = doi:10.1088/1126-6708/2008/11/043;%%
%
\bibitem{MPtop}
M.~Marino and P.~Putrov,
``Exact Results in ABJM Theory from Topological Strings,''
JHEP {\bf 1006}, 011 (2010)
%doi:10.1007/JHEP06(2010)011
[arXiv:0912.3074 [hep-th]].
%%CITATION = doi:10.1007/JHEP06(2010)011;%%
%
\bibitem{DT}
N.~Drukker and D.~Trancanelli,
``A Supermatrix model for N=6 super Chern-Simons-matter theory,''
JHEP {\bf 1002}, 058 (2010)
%doi:10.1007/JHEP02(2010)058
[arXiv:0912.3006 [hep-th]].
%%CITATION = doi:10.1007/JHEP02(2010)058;%%
%
\bibitem{KWY}
A.~Kapustin, B.~Willett and I.~Yaakov,
``Exact Results for Wilson Loops in Superconformal Chern-Simons Theories with Matter,''
JHEP {\bf 1003}, 089 (2010)
%doi:10.1007/JHEP03(2010)089
[arXiv:0909.4559 [hep-th]].
%%CITATION = doi:10.1007/JHEP03(2010)089;%%
%
\bibitem{KM}
N.~Kubo and S.~Moriyama,
``Two-Point Functions in ABJM Matrix Model,''
JHEP {\bf 1805}, 181 (2018)
%doi:10.1007/JHEP05(2018)181
[arXiv:1803.07161 [hep-th]].
%
\bibitem{MatsumotoM}
S.~Matsumoto and S.~Moriyama,
``ABJ Fractional Brane from ABJM Wilson Loop,''
JHEP {\bf 1403}, 079 (2014)
%doi:10.1007/JHEP03(2014)079
[arXiv:1310.8051 [hep-th]].
%%CITATION = doi:10.1007/JHEP03(2014)079;%%
%
\bibitem{HHMO}
Y.~Hatsuda, M.~Honda, S.~Moriyama and K.~Okuyama,
``ABJM Wilson Loops in Arbitrary Representations,''
JHEP {\bf 1310}, 168 (2013)
%doi:10.1007/JHEP10(2013)168
[arXiv:1306.4297 [hep-th]].
%%CITATION = doi:10.1007/JHEP10(2013)168;%%
%
\bibitem{MatsunoM}
S.~Matsuno and S.~Moriyama,
``Giambelli Identity in Super Chern-Simons Matrix Model,''
J.\ Math.\ Phys.\  {\bf 58}, no. 3, 032301 (2017)
%doi:10.1063/1.4978229
[arXiv:1603.04124 [hep-th]].
%%CITATION = doi:10.1063/1.4978229;%%
%
\bibitem{FM}
T.~Furukawa and S.~Moriyama,
``Jacobi-Trudi Identity in Super Chern-Simons Matrix Model,''
SIGMA {\bf 14}, 049 (2018)
%doi:10.3842/SIGMA.2018.049
[arXiv:1711.04893 [hep-th]].
%%CITATION = doi:10.3842/SIGMA.2018.049;%%
%
\bibitem{KOS}
A.~Kuniba, Y.~Ohta and J.~Suzuki,
``Quantum Jacobi-Trudi and Giambelli formulae for $U_q(B_r^{(1)})$ from analytic Bethe ansatz,''
J.\ Phys.\ A {\bf 28}, 6211 (1995)
%doi:10.1088/0305-4470/28/21/024
[hep-th/9506167].
%%CITATION = doi:10.1088/0305-4470/28/21/024;%%
%
\bibitem{HL}
J.~Harnad and E.~Lee,
``Symmetric polynomials, generalized Jacobi-Trudi identities and $\tau$-functions,''
J.\ Math.\ Phys.\  {\bf 59}, no.\ 9, 091411 (2018)
%doi:10.1063/1.5051546
[arXiv:1304.0020 [math-ph]].
%%CITATION = doi:10.1063/1.5051546;%%
%
\bibitem{AKLTZ}
A.~Alexandrov, V.~Kazakov, S.~Leurent, Z.~Tsuboi and A.~Zabrodin,
``Classical tau-function for quantum spin chains,''
JHEP {\bf 1309}, 064 (2013)
%doi:10.1007/JHEP09(2013)064
[arXiv:1112.3310 [math-ph]].
%%CITATION = doi:10.1007/JHEP09(2013)064;%%
%
\bibitem{AZ}
A.~Alexandrov and A.~Zabrodin,
``Free fermions and tau-functions,''
J.\ Geom.\ Phys.\  {\bf 67}, 37 (2013)
%doi:10.1016/j.geomphys.2013.01.007
[arXiv:1212.6049 [math-ph]].
%%CITATION = doi:10.1016/j.geomphys.2013.01.007;%%
%
\bibitem{md}
I.~G.~Macdonald,
``Symmetric functions and Hall polynomials,''
Oxford university press, (1998).
%
\bibitem{md9}
I.~G.~Macdonald,
``Schur functions: theme and variations,''
S\'eminaire Lotharingien de Combinatoire (Saint-Nabor, 1992) {\bf 498}, 5-39 (1992).
%
\bibitem{Moens}
E.~Moens,
``Supersymmetric Schur functions and Lie superalgebra representations,''
Doctoral dissertation, Ghent University (2007). 
%
\bibitem{YM}
N.~Beisert {\it et al.},
``Review of AdS/CFT Integrability: An Overview,''
Lett.\ Math.\ Phys.\  {\bf 99}, 3 (2012)
%doi:10.1007/s11005-011-0529-2
[arXiv:1012.3982 [hep-th]].
%%CITATION = doi:10.1007/s11005-011-0529-2;%%
%
\bibitem{MP}
M.~Marino and P.~Putrov,
``ABJM theory as a Fermi gas,''
J.\ Stat.\ Mech.\  {\bf 1203}, P03001 (2012)
%doi:10.1088/1742-5468/2012/03/P03001
[arXiv:1110.4066 [hep-th]].
%%CITATION = doi:10.1088/1742-5468/2012/03/P03001;%%
%
\bibitem{MN1}
S.~Moriyama and T.~Nosaka,
``Partition Functions of Superconformal Chern-Simons Theories from Fermi Gas Approach,''
JHEP {\bf 1411}, 164 (2014)
%doi:10.1007/JHEP11(2014)164
[arXiv:1407.4268 [hep-th]].
%%CITATION = doi:10.1007/JHEP11(2014)164;%%
%
\bibitem{EGSV}
J.~Escobedo, N.~Gromov, A.~Sever and P.~Vieira,
``Tailoring Three-Point Functions and Integrability,''
JHEP {\bf 1109}, 028 (2011)
%doi:10.1007/JHEP09(2011)028
[arXiv:1012.2475 [hep-th]].
%%CITATION = doi:10.1007/JHEP09(2011)028;%%
%
\bibitem{F}
O.~Foda,
``N=4 SYM structure constants as determinants,''
JHEP {\bf 1203}, 096 (2012)
%doi:10.1007/JHEP03(2012)096
[arXiv:1111.4663 [math-ph]].
%%CITATION = doi:10.1007/JHEP03(2012)096;%%
%
\bibitem{FS}
O.~Foda and G.~Schrader,
``XXZ scalar products, Miwa variables and discrete KP,''
New Trends in Quantum Integrable Systems 61-80,
Singapore, World Scientific (2010),
[arXiv:1003.2524 [math-ph]].
%
\bibitem{T}
K.~Takasaki,
``KP and Toda tau functions in Bethe ansatz,''
New Trends in Quantum Integrable Systems 373-392,
Singapore, World Scientific (2010),
%doi:10.1142/9789814324373_0019
[arXiv:1003.3071 [math-ph]].
%%CITATION = doi:10.1142/9789814324373_0019;%%
%
\bibitem{HaOk}
Y.~Hatsuda and K.~Okuyama,
``Exact results for ABJ Wilson loops and open-closed duality,''
JHEP {\bf 1610}, 132 (2016)
%doi:10.1007/JHEP10(2016)132
[arXiv:1603.06579 [hep-th]].
%%CITATION = doi:10.1007/JHEP10(2016)132;%%
%
\bibitem{KiyoshigeM}
K.~Kiyoshige and S.~Moriyama,
``Dualities in ABJM Matrix Model from Closed String Viewpoint,''
JHEP {\bf 1611}, 096 (2016)
%doi:10.1007/JHEP11(2016)096
[arXiv:1607.06414 [hep-th]].
%%CITATION = doi:10.1007/JHEP11(2016)096;%%
%
\bibitem{DMP1}
N.~Drukker, M.~Marino and P.~Putrov,
``From weak to strong coupling in ABJM theory,''
Commun.\ Math.\ Phys.\  {\bf 306}, 511 (2011)
%doi:10.1007/s00220-011-1253-6
[arXiv:1007.3837 [hep-th]].
%%CITATION = doi:10.1007/s00220-011-1253-6;%%
%
\bibitem{HKPT}
C.~P.~Herzog, I.~R.~Klebanov, S.~S.~Pufu and T.~Tesileanu,
``Multi-Matrix Models and Tri-Sasaki Einstein Spaces,''
Phys.\ Rev.\ D {\bf 83}, 046001 (2011)
%doi:10.1103/PhysRevD.83.046001
[arXiv:1011.5487 [hep-th]].
%%CITATION = doi:10.1103/PhysRevD.83.046001;%%
%
\bibitem{DMP2}
N.~Drukker, M.~Marino and P.~Putrov,
``Nonperturbative aspects of ABJM theory,''
JHEP {\bf 1111}, 141 (2011)
%doi:10.1007/JHEP11(2011)141
[arXiv:1103.4844 [hep-th]].
%%CITATION = doi:10.1007/JHEP11(2011)141;%%
%
\bibitem{FHM}
H.~Fuji, S.~Hirano and S.~Moriyama,
``Summing Up All Genus Free Energy of ABJM Matrix Model,''
JHEP {\bf 1108}, 001 (2011)
%doi:10.1007/JHEP08(2011)001
[arXiv:1106.4631 [hep-th]].
%%CITATION = doi:10.1007/JHEP08(2011)001;%%
%
\bibitem{KEK}
M.~Hanada, M.~Honda, Y.~Honma, J.~Nishimura, S.~Shiba and Y.~Yoshida,
``Numerical studies of the ABJM theory for arbitrary N at arbitrary coupling constant,''
JHEP {\bf 1205}, 121 (2012)
%doi:10.1007/JHEP05(2012)121
[arXiv:1202.5300 [hep-th]].
%%CITATION = doi:10.1007/JHEP05(2012)121;%%
%
\bibitem{HMO2}
Y.~Hatsuda, S.~Moriyama and K.~Okuyama,
``Instanton Effects in ABJM Theory from Fermi Gas Approach,''
JHEP {\bf 1301}, 158 (2013)
%doi:10.1007/JHEP01(2013)158
[arXiv:1211.1251 [hep-th]].
%%CITATION = doi:10.1007/JHEP01(2013)158;%%
%
\bibitem{CM}
F.~Calvo and M.~Marino,
``Membrane instantons from a semiclassical TBA,''
JHEP {\bf 1305}, 006 (2013)
%doi:10.1007/JHEP05(2013)006
[arXiv:1212.5118 [hep-th]].
%%CITATION = doi:10.1007/JHEP05(2013)006;%%
%
\bibitem{HMO3}
Y.~Hatsuda, S.~Moriyama and K.~Okuyama,
``Instanton Bound States in ABJM Theory,''
JHEP {\bf 1305}, 054 (2013)
%doi:10.1007/JHEP05(2013)054
[arXiv:1301.5184 [hep-th]].
%%CITATION = doi:10.1007/JHEP05(2013)054;%%
%
\bibitem{HMMO}
Y.~Hatsuda, M.~Marino, S.~Moriyama and K.~Okuyama,
``Non-perturbative effects and the refined topological string,''
JHEP {\bf 1409}, 168 (2014)
%doi:10.1007/JHEP09(2014)168
[arXiv:1306.1734 [hep-th]].
%%CITATION = doi:10.1007/JHEP09(2014)168;%%
%
\bibitem{HO}
M.~Honda and K.~Okuyama,
``Exact results on ABJ theory and the refined topological string,''
JHEP {\bf 1408}, 148 (2014)
%doi:10.1007/JHEP08(2014)148
[arXiv:1405.3653 [hep-th]].
%%CITATION = doi:10.1007/JHEP08(2014)148;%%
%
\bibitem{HMO1}
Y.~Hatsuda, S.~Moriyama and K.~Okuyama,
``Exact Results on the ABJM Fermi Gas,''
JHEP {\bf 1210}, 020 (2012)
%doi:10.1007/JHEP10(2012)020
[arXiv:1207.4283 [hep-th]].
%%CITATION = doi:10.1007/JHEP10(2012)020;%%
%
\bibitem{PY}
P.~Putrov and M.~Yamazaki,
``Exact ABJM Partition Function from TBA,''
Mod.\ Phys.\ Lett.\ A {\bf 27}, 1250200 (2012)
%doi:10.1142/S0217732312502008
[arXiv:1207.5066 [hep-th]].
%%CITATION = doi:10.1142/S0217732312502008;%%
%
\bibitem{TW1}
C.~A.~Tracy and H.~Widom,
``Fredholm determinants and the mKdV/sinh-Gordon hierarchies,''
Commun.\ Math.\ Phys.\  {\bf 179}, 1 (1996)
[solv-int/9506006].
%
\bibitem{TW2}
C.~A.~Tracy and H.~Widom,
``Proofs of two conjectures related to the thermodynamic Bethe ansatz,''
Commun.\ Math.\ Phys.\  {\bf 179}, 667 (1996)
%doi:10.1007/BF02100102
[solv-int/9509003].
%%CITATION = doi:10.1007/BF02100102;%%
%
\bibitem{Z}
A.~B.~Zamolodchikov,
``Painleve III and 2-d polymers,''
Nucl.\ Phys.\ B {\bf 432}, 427 (1994)
%doi:10.1016/0550-3213(94)90029-9
[hep-th/9409108].
%%CITATION = doi:10.1016/0550-3213(94)90029-9;%%
%
\bibitem{GHM1}
A.~Grassi, Y.~Hatsuda and M.~Marino,
``Topological Strings from Quantum Mechanics,''
Annales Henri Poincare {\bf 17}, no. 11, 3177 (2016)
%doi:10.1007/s00023-016-0479-4
[arXiv:1410.3382 [hep-th]].
%%CITATION = doi:10.1007/s00023-016-0479-4;%%
%
\bibitem{KaMa}
R.~Kashaev and M.~Marino,
``Operators from mirror curves and the quantum dilogarithm,''
Commun.\ Math.\ Phys.\  {\bf 346}, no. 3, 967 (2016)
%doi:10.1007/s00220-015-2499-1
[arXiv:1501.01014 [hep-th]].
%%CITATION = doi:10.1007/s00220-015-2499-1;%%
%
\bibitem{MZ}
M.~Marino and S.~Zakany,
``Matrix models from operators and topological strings,''
Annales Henri Poincare {\bf 17}, no. 5, 1075 (2016)
%doi:10.1007/s00023-015-0422-0
[arXiv:1502.02958 [hep-th]].
%%CITATION = doi:10.1007/s00023-015-0422-0;%%
%
\bibitem{KMZ}
R.~Kashaev, M.~Marino and S.~Zakany,
``Matrix models from operators and topological strings, 2,''
Annales Henri Poincare {\bf 17}, no. 10, 2741 (2016)
%doi:10.1007/s00023-016-0471-z
[arXiv:1505.02243 [hep-th]].
%%CITATION = doi:10.1007/s00023-016-0471-z;%%
%
\bibitem{WZH}
X.~Wang, G.~Zhang and M.~x.~Huang,
``New Exact Quantization Condition for Toric Calabi-Yau Geometries,''
Phys.\ Rev.\ Lett.\  {\bf 115}, 121601 (2015)
%doi:10.1103/PhysRevLett.115.121601
[arXiv:1505.05360 [hep-th]].
%%CITATION = doi:10.1103/PhysRevLett.115.121601;%%
%
\bibitem{GKMR}
J.~Gu, A.~Klemm, M.~Marino and J.~Reuter,
``Exact solutions to quantum spectral curves by topological string theory,''
JHEP {\bf 1510}, 025 (2015)
%doi:10.1007/JHEP10(2015)025
[arXiv:1506.09176 [hep-th]].
%%CITATION = doi:10.1007/JHEP10(2015)025;%%
%
\bibitem{CGrM}
S.~Codesido, A.~Grassi and M.~Marino,
``Spectral Theory and Mirror Curves of Higher Genus,''
Annales Henri Poincare {\bf 18}, no. 2, 559 (2017)
%doi:10.1007/s00023-016-0525-2
[arXiv:1507.02096 [hep-th]].
%%CITATION = doi:10.1007/s00023-016-0525-2;%%
%
\bibitem{SWH}
K.~Sun, X.~Wang and M.~x.~Huang,
``Exact Quantization Conditions, Toric Calabi-Yau and Nonperturbative Topological String,''
JHEP {\bf 1701}, 061 (2017)
%doi:10.1007/JHEP01(2017)061
[arXiv:1606.07330 [hep-th]].
%%CITATION = doi:10.1007/JHEP01(2017)061;%%
%
\bibitem{CGuM}
S.~Codesido, J.~Gu and M.~Marino,
``Operators and higher genus mirror curves,''
JHEP {\bf 1702}, 092 (2017)
%doi:10.1007/JHEP02(2017)092
[arXiv:1609.00708 [hep-th]].
%%CITATION = doi:10.1007/JHEP02(2017)092;%%
%
\bibitem{MN3}
S.~Moriyama and T.~Nosaka,
``Exact Instanton Expansion of Superconformal Chern-Simons Theories from Topological Strings,''
JHEP {\bf 1505}, 022 (2015)
%doi:10.1007/JHEP05(2015)022
[arXiv:1412.6243 [hep-th]].
%%CITATION = doi:10.1007/JHEP05(2015)022;%%
%
\bibitem{HHO}
Y.~Hatsuda, M.~Honda and K.~Okuyama,
``Large N non-perturbative effects in $\mathcal{N}=4$ superconformal Chern-Simons theories,''
JHEP {\bf 1509}, 046 (2015)
%doi:10.1007/JHEP09(2015)046
[arXiv:1505.07120 [hep-th]].
%%CITATION = doi:10.1007/JHEP09(2015)046;%%
%
\bibitem{MNN}
S.~Moriyama, S.~Nakayama and T.~Nosaka,
``Instanton Effects in Rank Deformed Superconformal Chern-Simons Theories from Topological Strings,''
JHEP {\bf 1708}, 003 (2017)
%doi:10.1007/JHEP08(2017)003
[arXiv:1704.04358 [hep-th]].
%%CITATION = doi:10.1007/JHEP08(2017)003;%%
%
\bibitem{MNY}
S.~Moriyama, T.~Nosaka and K.~Yano,
``Superconformal Chern-Simons Theories from del Pezzo Geometries,''
JHEP {\bf 1711}, 089 (2017)
%doi:10.1007/JHEP11(2017)089
[arXiv:1707.02420 [hep-th]].
%%CITATION = doi:10.1007/JHEP11(2017)089;%%
%
\bibitem{KMN}
N.~Kubo, S.~Moriyama and T.~Nosaka,
``Symmetry Breaking in Quantum Curves and Super Chern-Simons Matrix Models,''
arXiv:1811.06048 [hep-th].
%%CITATION = ARXIV:1811.06048;%%
%
\bibitem{GHM2}
A.~Grassi, Y.~Hatsuda and M.~Marino,
``Quantization conditions and functional equations in ABJ(M) theories,''
J.\ Phys.\ A {\bf 49}, no. 11, 115401 (2016)
%doi:10.1088/1751-8113/49/11/115401
[arXiv:1410.7658 [hep-th]].
%%CITATION = doi:10.1088/1751-8113/49/11/115401;%%
%
\bibitem{H}
M.~Honda,
``Exact relations between M2-brane theories with and without Orientifolds,''
JHEP {\bf 1606} (2016) 123
%doi:10.1007/JHEP06(2016)123
[arXiv:1512.04335 [hep-th]].
%%CITATION = doi:10.1007/JHEP06(2016)123;%%
%
\bibitem{MS2}
S.~Moriyama and T.~Suyama,
``Orthosymplectic Chern-Simons Matrix Model and Chirality Projection,''
JHEP {\bf 1604} (2016) 132
%doi:10.1007/JHEP04(2016)132
[arXiv:1601.03846 [hep-th]].
%%CITATION = doi:10.1007/JHEP04(2016)132;%%
%
\bibitem{MN5}
S.~Moriyama and T.~Nosaka,
``Orientifold ABJM Matrix Model: Chiral Projections and Worldsheet Instantons,''
JHEP {\bf 1606} (2016) 068
%doi:10.1007/JHEP06(2016)068
[arXiv:1603.00615 [hep-th]].
%%CITATION = doi:10.1007/JHEP06(2016)068;%%
%
\bibitem{BGT}
G.~Bonelli, A.~Grassi and A.~Tanzini,
``Quantum curves and $q$-deformed Painlev\'e equations,''
arXiv:1710.11603 {[}hep-th{]}.
%%CITATION = ARXIV:1710.11603;%%
%
\bibitem{KNY}
K.~Kajiwara, M.~Noumi and Y.~Yamada,
``Geometric Aspects of Painlev\'e Equations,''
J.~Phys.~A: Math.~Theor. {\bf 50}, 073001 (2017)
[arxiv:1509.08186 [nlin]].
%
\bibitem{ZJ}
P.~Zinn-Justin,
``HCIZ integral and 2-D Toda lattice hierarchy,''
Nucl.\ Phys.\ B {\bf 634}, 417 (2002)
%doi:10.1016/S0550-3213(02)00374-7
[math-ph/0202045].
%%CITATION = doi:10.1016/S0550-3213(02)00374-7;%%
%
\bibitem{ADKMV}
M.~Aganagic, R.~Dijkgraaf, A.~Klemm, M.~Marino and C.~Vafa,
``Topological strings and integrable hierarchies,''
Commun.\ Math.\ Phys.\  {\bf 261}, 451 (2006)
%doi:10.1007/s00220-005-1448-9
[hep-th/0312085].
%%CITATION = doi:10.1007/s00220-005-1448-9;%%
%
\bibitem{DV}
R.~Dijkgraaf and C.~Vafa,
``Toda Theories, Matrix Models, Topological Strings, and N=2 Gauge Systems,''
arXiv:0909.2453 [hep-th].
%%CITATION = ARXIV:0909.2453;%%
%
\end{thebibliography}
\end{document}